\documentclass[a4paper,english]{article}
\usepackage[T1]{fontenc}
\usepackage{RRA4}
\RRNo{6474}
\usepackage{hyperref}
\usepackage{algorithmic}
\usepackage{algorithm}
\usepackage{epsfig}
\usepackage{amssymb}
\usepackage{amsmath}
\usepackage{amsfonts}

\usepackage{a4wide}
\usepackage{epsfig}
\usepackage{enumerate}
\usepackage{amsmath}

\usepackage{amsthm}
\theoremstyle{plain}

\theoremstyle{definition}

\theoremstyle{remark}

\usepackage{babel}

\RRdate{February 2008}
\RRauthor{
Luca Muscariello \thanks{luca.muscariello@orange-ftgroup.com}  \and 
Diego Perino\thanks{diego.perino@orange-ftgroup.com} 
}
\authorhead{Luca Muscariello and Diego Perino}
\RRtitle{ Ing\`enierie du Traffic par Flot\\ Une Approche Multi-chemins Pratique}
\RRetitle{Early Experiences in Traffic Engineering Exploiting Path Diversity: A Practical Approach}
\titlehead{Flow-Aware Traffic Engineering}
\RRresume{Les algorithmes de routage dynamique poss\`edent de solides fondements th\'eoriques qui les rendent aptes \'a une impl\'ementation r\`eelle dans diff\`erents r\`eseaux op\`erationnels. Ces algorithmes poss\`edent de nombreuses propri\`et\`es propres aux contr\^oles de congestion gr\^ace \'a l'utilisation de m\'ecanismes de signalisation explicite et de contr\^ole des flots \'a la source.

Dans cet article, nous proposons une formulation lin\`eaire du probl\`eme de multiflot avec contr\^ole de congestion et crit\`ere d'\`equit\`e de type "max-min". Les performances obtenues par l'exploitation de chemins multiples sont encourageants, que cela soit en routage intra-domaine ou dans les r\`eseaux mesh sans fil, et incitent \'a une impl\`ementation r\`eelle, en particulier dans le cas de matrices de trafic non-uniformes et de "points chauds" de demandes.

Nous proposons une approche par flots qui int\`egre naturellement les multi-chemins aux m\`ecanismes actuels de contr\^ole de congestion, et nous l'\`evaluons par rapport aux solutions actuelles. Les architectures par flot sont r\`ealisables, robustes, capables de passer \'a l'\`echelle, et quasi-optimales lorsque les flots ont un d\`ebit-cr\^ete explicite. Dans un contexte r\`ealiste, notre solution poss\`ede donc les propri\`et\`es d'une solution optimale ainsi que les avantages de l'approche par flots. }

\RRabstract{Recent literature has proved that stable dynamic routing algorithms
have solid theoretical foundation that makes them suitable to be implemented 
in a real protocol, and used in practice in many different operational network contexts.
Such algorithms inherit much of the properties of congestion controllers implementing 
one of the possible combination of AQM/ECN schemes at nodes and 
flow control at sources. 

In this paper we propose a linear program formulation of the multi-commodity flow problem
with congestion control, under max-min fairness, comprising demands with or without exogenous 
peak rates. Our evaluations of the gain, using path diversity, in scenarios as intra-domain traffic engineering 
and wireless mesh networks encourages real implementations, especially in presence of hot spots demands
and non uniform traffic matrices.

We propose a flow aware perspective of the subject by using 
a natural multi-path extension to current congestion controllers
and show its performance with respect to current proposals.
Since flow aware architectures exploiting path diversity are feasible, 
scalable, robust and nearly optimal in presence of flows with exogenous
peak rates, we claim that our solution rethinked in the context of
realistic traffic assumptions performs as better as an optimal approach
with all the additional benefits of the flow aware paradigm.}
\RRmotcle{Multi-chemins, Contr\^ole de congestion, Ing\`enierie du traffic, Architecture par flots}
\RRkeyword{Multi-path Routing,  Congestion control, Traffic Engineering, Flow-Aware Architectures}
\RRprojets{Gang}
\RRtheme{\THCom}
\URRocq  
\begin{document}
\makeRR   
\section{Introduction}
Traffic engineering is usually perceived as an off-line functionality in order to improve performance 
by better matching network resources to traffic demands. Optimisation is performed off-line as 
demands are averages over the long term that do not consider traffic 
fluctuation over smaller time scales. Intra-domain routing optimisation, by means of OSPF cost parametrisation
\cite{Thorup00,Thorup02}, is typical example of the aforementioned problem. 
The max degree of responsiveness is guaranteed at the long term, in a daily or weekly basis
for instance. The objective of ISPs is to get rid of a given traffic matrix at the minimum cost, which
is estimated as global expenditures for link upgrades. Therefore minimising the maximum link load is
a natural objective.
Such an engineered network is not robust to any finer grained traffic fluctuation, as large number of failures, traffic
flash crowds, BGP re-routes, application re-routing. In current backbones these effects are mitigated by over-provisioned
links.
In other contexts, of increasing importance nowadays as wireless mesh networks, scarceness of resources 
push network engineers to accept a higher degree of responsiveness in order to exploit any single piece
of unused capacity.  We only consider wireless mesh resulting from a radio engineered network, 
where links guarantee a minimum availability.

Literature on optimal routing in networks starts from seminal works as
 \cite{Bertsekas84,Bertsekas87}  or \cite{Gerla74,Gallager77}. This comprises centralised and 
decentralised strategies
proposed in the very beginning of the ARPAnet project. Historically the main outcome of optimal 
routing has been shortest-path or, at most, minimum-cost routing. 
Early experiences on dynamic routing \cite{Crowcroft} have kept it back in spite of its potential,
as sensitivity to congestion has long been refused for being complex, unstable, not prone to be 
easily deployed.

There is a quite large recent literature on multi-path routing, raised from the need to develop dynamic yet stable 
algorithms, sensitive to congestion for many applications that could effectively exploit unused resources
within the network.

In the Internet, we believe that very specific applications can tolerate such dynamic environment.
Adaptive video streaming is, probably, the main application that would definitely be able to tolerate variability
and, at the same time, take advantage of unused capacity.
It is worth recalling that video applications are going to be the main part of Internet traffic very soon.
Video streams usually last very long, and will probably last longer as better applications and contents
will be available.  Furthermore it is likely that such contents will be transported by a number of non-specified protocols, 
subject to non better specified fairness criteria.

Conversational applications should be kept away from being routed over multiple routes but also other
data services that are made of flows that last potentially very short. Also, much of the present data traffic 
if it includes mail, web, instant messaging.
However, P2P file sharing applications or content delivery networks (CDN) are prone to well exploit path diversity as they
intrinsically are robust to rate fluctuations.
Other data applications are downloads of software updates even though they can be included in the class
of P2P file sharing systems.

In the last few years, intense research on multi-path routing has progressed.
Theory of optimisation has been applied to develop distributed algorithms solving
a global optimisation problem, see \cite{Srikant06, Rexford07, RexfordCo07,Low99,Kelly05,Paganini07,Voice07,
Shroff06}. Optimisation explicits the problem of resource allocation under a chosen fairness criteria.
Control theory has been used to obtain delay stability of distributed optimal schemes, see 
\cite{Srikant06,katabi05,Kelly05,Voice07}.
Other research considers dynamic flow level models \cite{Key06,Key07,KeyInf07,oueslati07}
in order to take into account arrivals and departures of user's sessions.

In  this paper we support the deployment of a flow aware architecture exploiting path diversity for
a specific class of applications, rate adaptive video streaming for instance or CDN
and P2P file sharing. In such set up we show that flow aware paradigm is nearly optimal  for any 
tipical traffic demand requiring the use of 
multiple paths without the need to assume any kind of common transport protocol among users, any common fairness semantic 
and any kind of cooperation between  users, and network nodes as well.

In Section \ref{sec:traffic} we explain our definition of network flows while in
Section \ref{sec:model} the modelling framework for optimal routing and 
congestion control is introduced. The section includes 
a set of examples on toy networks. 

Finally we introduce our main outcome in term of performance evaluation
of the optimal solution of large problems, with an original
linear program formulation under max-min fairness that is used to 
evaluate large problems in Section \ref{sec:largelp}.

Section \ref{sec:arcprot} introduces our main original outcome as a new multi-path congestion
controller called MIRTO. The algorithm is born 
inferring an optimal strategy from  previous sections.
Moreover this algorithm is evaluated within the framework of a flow aware architecture,
bringing new arguments in favour of such network paradigm.

\section{Traffic Characteristics}\label{sec:traffic}
IP traffic on a network link can be considered as a superposition of 
independent sessions, each session relating to some piece of user activity 
and being manifested by the transmission of a collection of flows.

Sessions and flows are defined locally at a considered network interface. 
Flows can generally be identified by common values in packet header fields 
(e.g., the 5-tuple of IP addresses, port numbers and transport protocol) 
and the fact that the interval between such packets is less than some time 
out value (20s, say). It is not usually possible to identify sessions just 
from data in packets and this notion cannot therefore be used for resource 
allocation. Nevertheless we are more inclined to think about user sessions
than protocol defined flows.

A more significant flow characteristic is the exogenous peak rate at which 
a flow can be emitted. This is the highest rate the flow would attain if 
the link were of unlimited capacity. This limit may be due to the user access 
capacity, the maximum TCP receive window,  or the current available bandwidth on other 
links of the path, or the stream rate in case of video applications for instance.

In the rest of the paper we will use interchangeably the terms demand and flow.

\section{Modelling framework} \label{sec:model}
The network topology is modelled by a connected graph $G=(N,L)$ given as a set of nodes and links. 
Let $A = [a_{ij}]$  the adjacency matrix,
$a_{ij} = 1$ if there exists a directional link between $i$ and $j$ and $a_{ij} = 0$ otherwise.

The network carries traffic generated by a set of demands $\Gamma$, each demand $d$  is given with a triple $(s^d,e^d,p^d)$, with $s\in \mathcal{S}$,
$e\in \mathcal{E}$, source and destination nodes with $\mathcal{S},\mathcal{E}\subseteq N$, and $p\in \mathbb{R^+}$ exogenous peak rate. In our model a network flow $d$ gets a share $x_{ij}^d$ of the capacity $c_{ij}$ at each link $0 \leq x_{ij}^d\leq \min(c_{ij},p)$. 
$x_{ij}^d(t)$ is a fluid approximation of the rate at which the source $d$ is sending at time $t$ through link $ij$.

The network flow can be slit among different paths that are made available by a network 
protocol at an ingress node.  We make no modelling assumption whether paths are disjoint, 
however the ability to create more path diversity helps design highly robust network 
routing protocols.

In paragraph \ref{subsec:optim} we model route selection and bandwidth sharing as an optimisation problem 
that maximises user satisfaction and minimise network congestion under a specified fairness criteria.

\begin{table}[ht]
		\centering
		\begin{tabular}{|l|l|}
		\hline
		\textbf{Symbol} & \textbf{Meaning}\\
		\hline
		$N$ & node set\\ 
		\hline
		$L$ & link set\\
		\hline
		$\Gamma$ & demand set\\
		\hline
		$d$ & demand number\\
		\hline
		$\mathcal{S}$ & source set\\
		\hline
		$\mathcal{E}$ & destination set\\
		\hline
		$s^d$ & demand $d$ source node \\
        \hline
		$e^d$ & demand $d$ destination node \\
		\hline
		$p^d$ & demand $d$ exogenous peak rate \\
    	\hline
    	$P^d$ & path set of demand $d$\\
		\hline
		$k$ & path number\\
    	\hline
		$L_k^d$ & link set of demand $d$ over its $k^{th}$ path\\
    	\hline
		$C_{ij}$ & capacity of link $(i,j)$\\
    	\hline
		$x_{ij}^d$ & rate of demand $d$ over link $(i,j)$\\
		\hline
		$x_{k}^d$ & rate of demand $d$ over its $k^{th}$ path\\
		\hline
		$x^d$ & rate of demand $d$ ($\sum_{k}{x_k^d}$)\\
		\hline
		$\rho_{ij}$ & load on link $(i,j)$ ($\sum_{d \in \Gamma}\frac{x_{ij}^d}{C_{ij}}$)\\
		\hline		
		\end{tabular}
\caption{\label{tab:notation}Summary of notation used}
\end{table}

\subsection{Minimum cost routing}
The ability to create the set of optimal paths at the ingress of the network and make them available to the 
routing protocol requires a certain knowledge of the network status, as link load, path delay and length. 
However, as this can be done in practice by disseminating local measures,
the protocol must be also robust to state inaccuracy. 
Assuming perfect knowledge of network state, optimal routing can be formulated through the following  non linear
optimisation problem with linear constrained.
\begin{align}
 \mbox{minimise}  \sum_{i,j\in N} C\left(\frac{\sum_{d\in \Gamma} x_{ij}^d }{c_{ij}}\right) \nonumber
\end{align}
$\mbox{subject to}$
\begin{align}
 \sum_{k\in N} a_{ik}x_{ki}^d - \sum_{j \in N}a_{ji} x_{ij}^d = \left\{ \begin{array}{ll} p_i^d & \mbox{if}~~ i\in \mathcal{S}\\ -p_i^d & \mbox{if}~~ i\in \mathcal{E}\\ 0 & \mbox{otherwise}\end{array}\right. & \forall d\in \Gamma \label{eqn:flow}
\end{align}
constraint (\ref{eqn:flow}) models zero net flow for relay nodes, positive for source nodes and 
negative for destination nodes. This allows to obtain optimal routes directly from the optimisation
problem. $C$ can be thought modelling the link delay often used in traffic engineering formulations 
of the multi-commodity flow problem,

\begin{equation}\label{eqn:cost}
C(x_{ij}) = \frac{x_{ij}}{c_{ij}-x_{ij}}
\end{equation}

with this formula the cost function becomes the average delay in a M/M/1 queue as a consequence of the 
Kleinrock independence approximation and Jackson's Theorem.  This problem formulation dates back to 
\cite{Gerla74, Gallager77} in the context of minimum delay routing.
Using standard techniques in convex constrained optimisation (convex optimisation over a simplex ) in 
\cite{Bertsekas84, Bertsekas87}, it is shown that the optimal solution always select paths with minimum 
(and equal) firs cost derivatives for any strictly convex cost function. Therefore the problem can be re-formulated 
as a shortest path problem where path lengths are the first derivatives of the cost function along the path,
that can be written with abuse of notation, 

\begin{equation} \label{eqn:cost2}
C^{'}(x_p^d)=C^{'} \left(\frac{\sum_{d\in \Gamma} x_{ij}^d }{c_{ij}}\right)
\end{equation}

where  $x_p^d$ is the portion of flow of demand $d$ flowing through path $p$. This is what, 
in \cite{Bertsekas87}, Bertsekas and Gallager call first derivative path lengths. Therefore, at optimum 
all paths have equal lengths.  This fact will be use in the following section repeatedly.

Another plausible objective is to minimise the most loaded link, frequent in traffic engineering network 
operator's backbone optimization in conjunction with link capacity over-provisioning.
In the context of multi-path routing this has been used to design TEXCP \cite{katabi05}.

\subsection{Bandwidth sharing and fairness}
Bandwidth is shared  between flows according to a certain objective realised by 
one  transport protocol as TCP for data transfers making use of one of its congestion control protocols 
(Reno, Vegas, Cubic, high speed etc.) or TCP friendly rate control (TFRC) for adaptive streaming applications. 
In general these protocols realise different fairness criteria,  whilst in this context we assume that
all flows are subject to a common fairness objective. The problem formulation dates back to Kelly 
\cite{Kelly97}  where this problem is  formulated as  a non linear optimization problem with linear constrained 
with objective given by a utility function $U(x)$ of the flow rate $x$.
\begin{align}
 \mbox{maximise} \sum_{i\in \mathcal{S}, d\in \Gamma} U_d(\phi_i^d)  \nonumber
\end{align}
$\mbox{subject to}$
\begin{align}
 \sum_{k\in N} a_{ik}x_{ki}^d - \sum_{j \in N}a_{ji} x_{ij}^d = \left\{ \begin{array}{ll} \phi_i^d & \mbox{if}~~ i\in \mathcal{S}\\ -\phi_i^d & \mbox{if}~~ i\in \mathcal{E}\\ 0 & \mbox{otherwise}\end{array}\right. & \forall d\in \Gamma \label{eqn:flow}
\end{align}
\begin{align}
 \sum_{d\in \Gamma} x_{ij}^d \leq c_{ij} & \forall i,j\in L & 
\end{align}
demands are assumed elastic, meaning that they could get as much bandwidth as network
status allows, and have no exogenous peak rates.
A general class of utility functions has been introduces in \cite{Mo00},
\begin{equation} \label{eqn:alphafair}
 U_{d}(x) =  \left\{ 
\begin{array}{ll} w_d\log x & \alpha=1 \\ w_d(1-\alpha)^{-1}x^{1-\alpha} & \alpha \neq 1 \end{array} \right.
\end{equation}
if $\alpha \rightarrow \infty$ fairness criteria is max-min.\\
This formulation is widely  and succesfully used in network modeling. 
 \cite{Srikant06, Rexford07,Kelly98, Kelly03, Kelly05,Srikant04}  have 
considered the problem of single and multiple path routing and congestion control under this framework
as constraint (\ref{eqn:flow}) may count either a single or a multiple set of routes.

\subsection{User utility and network cost} \label{subsec:optim}
User utility and network cost are two conflicting objective in a mathematical 
formulation. \cite{ Rexford07, RexfordCo07,katabi05} have used the cost function as TE objective
likely modeled by the ISPs in ordered to keep low link loads, i.e. minimise costs
for upgrades. \cite{Srikant06,Low99,Kelly05,Paganini07,Voice07,Shroff06}
have just used the network cost as penalty function in place of hard constraints in
the optimisation framework.

Congestion sensitive multiple routes selection can be formulated as a mathematical program with non linear
objective and linear constraints. 
\begin{align}\label{eqn:opt}
 \mbox{maximise} \sum_{i\in \mathcal{S}, d\in \Gamma} U_d(\phi_i^d) - \sum_{i,j\in N} C\left(\frac{\sum_{d\in \Gamma} x_{ij}^d }{c_{ij}}\right) \nonumber
\end{align}
$\mbox{subject to}$
\begin{align}
 \sum_{k\in N} a_{ik}x_{ki}^d - \sum_{j \in N}a_{ji} x_{ij}^d = \left\{ \begin{array}{ll} \phi_i^d & \mbox{if}~~ i\in \mathcal{S}\\ -\phi_i^d & \mbox{if}~~ i\in \mathcal{E}\\ 0 & \mbox{otherwise}\end{array}\right. & \forall d\in \Gamma 
\end{align}
\begin{align}
 \sum_{d\in \Gamma} x_{ij}^d \leq c_{ij} & \forall i,j\in L & 
\end{align}
\begin{align}
\phi_i^d \leq p_d & ~~\forall d\in \Gamma ~~ \forall i\in \mathcal{S}
\end{align}
As a new additional constraints we add exgenous rates as this has significant impact in the
process of selection of optimal routes.

In this formulation the user utility is a function of the total flow rate traversing the network through the available
paths. In a path-demand formulation the flow rate$\phi$   of a given user  can be re-written as the sum of the
rates over the set of available paths $\mathcal{P}$, i.e. $\phi=\sum_{p\in\mathcal{P}} \phi_p$.
A user is free to coordinate sending rates over the paths jointly, aiming at maximise its own utility.
Consider now the folllowing relaxation of the objective
\begin{eqnarray} \label{eqn:opt2}
U(\sum_{p\in\mathcal{P}} \phi_p) \geq \sum_{p\in\mathcal{P}}U( \phi_p) 
\end{eqnarray}
each user's path would be seen as independent, in other words as if it were a separate user and 
the fairness objective would be at a path, and not user base.
A protocol designed observing such rule would break path coordination, while a network
imposing per link fair bandwith sharing, would realise this objective for any multi-path controller
regardless its original design.

\subsection{Toy Examples}
\begin{figure}[htb]
\begin{center} \label{fig:simple-topology}
$\begin{array}{cc}
\includegraphics[width=0.2\textwidth]{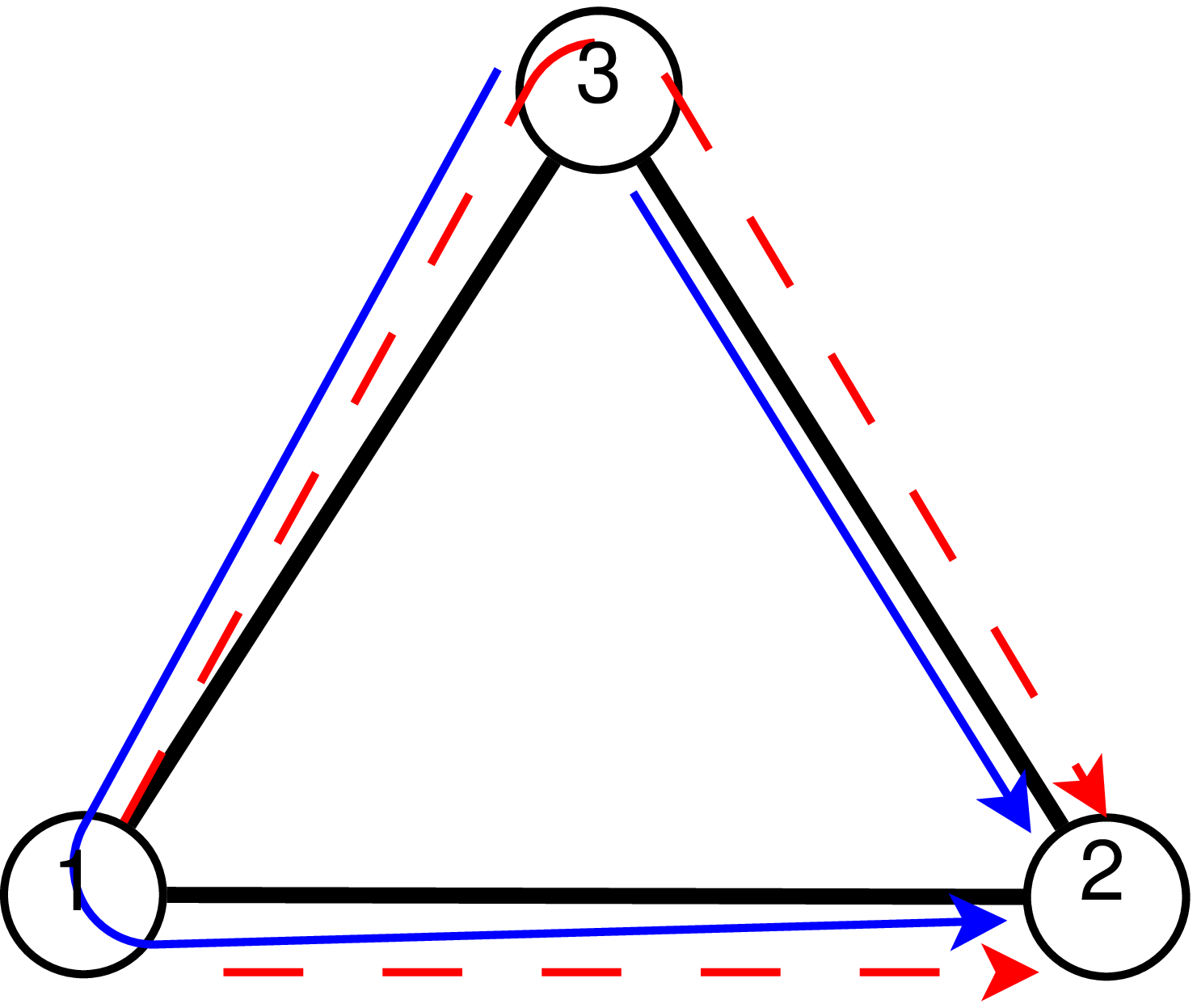} &
\includegraphics[width=0.2\textwidth]{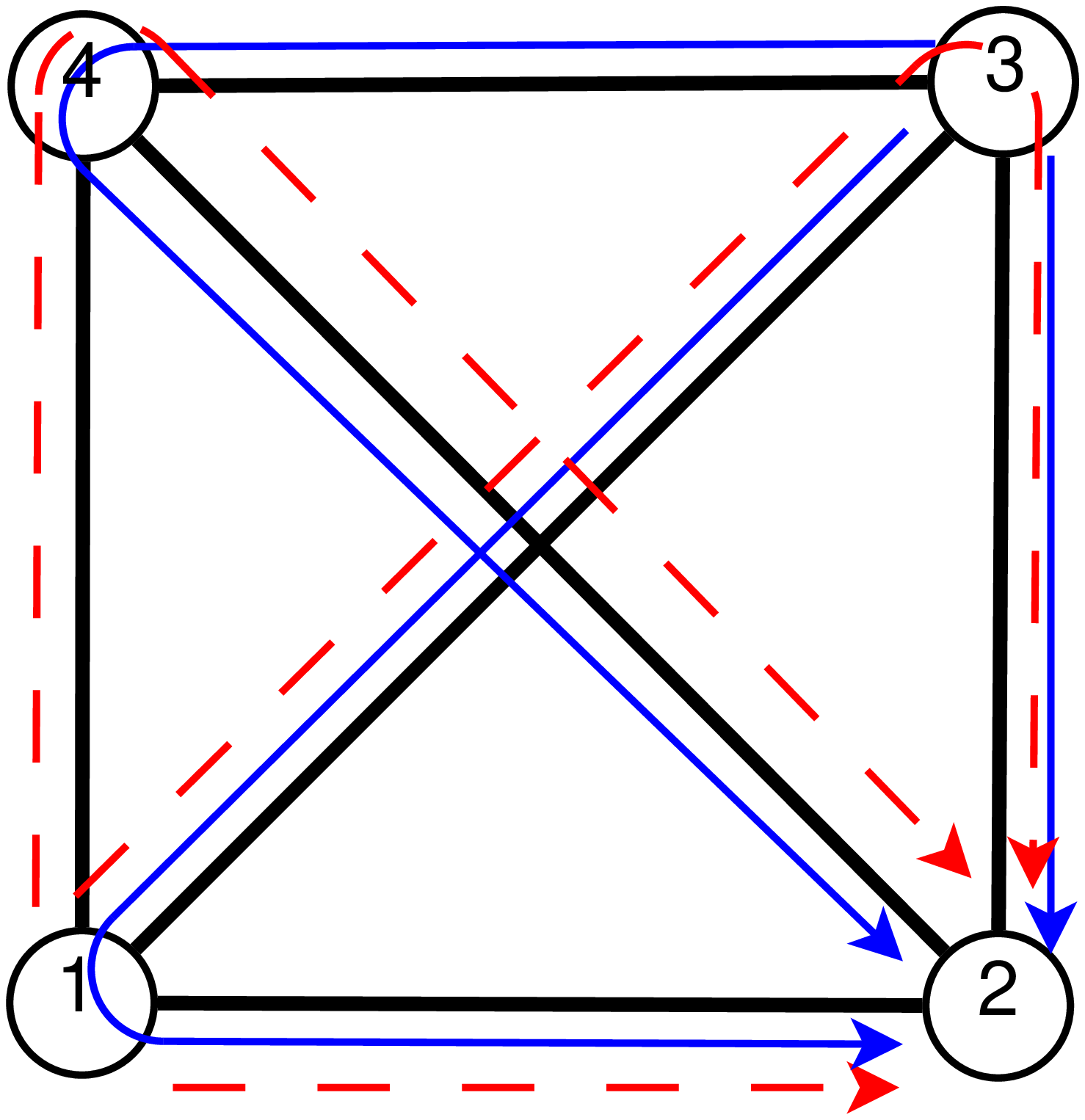}\\
\mbox{{\bf{(a)}}} & \mbox{{\bf{(b)}}}
\end{array}$
\end{center}
\caption{Full mesh triangle and square topologies. Hot-spot destination in node 2 and sources in node 1 and 3.	}
\end{figure}

In this section we consider two simple network topologies, a triangle and a square as depicted in 
fig.\ref{fig:simple-topology} with all available paths. All links are bi-derctional with the same capacity
$C$. Capacity $C_{12}=C_{31}$ is increased from $C$ to $15\times C$. 
For both scenarios nodes 1 and 3 send data to a single destination node number 2.
We find the global optimum of problem (\ref{eqn:opt}), using utility function (\ref{eqn:alphafair}) with $\alpha=2$
and cost function (\ref{eqn:cost}). We assume demands fully elastic.

\subsubsection{Triangle}
\label{sec:triangolo}
\begin{figure}[htb] 
\begin{center}
$\begin{array}{cc}
\includegraphics[width=0.45\textwidth]{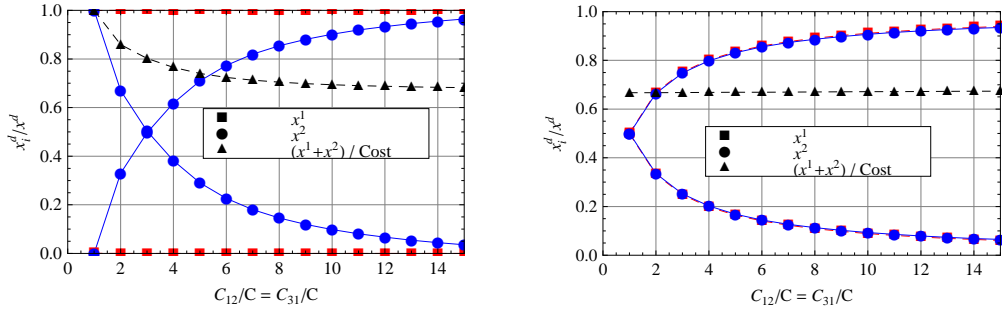} &
\includegraphics[width=0.45\textwidth]{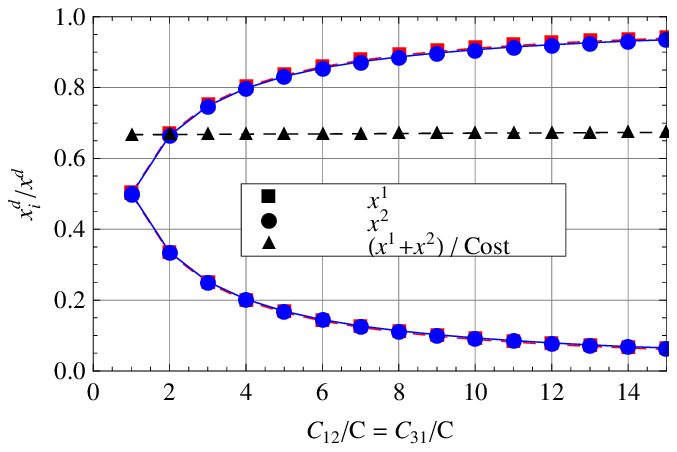} 
\end{array}$
\end{center}
\caption{Rate distribution among the available paths and user rate to network cost ratio.	}
\label{fig:triangolo-curves}
\end{figure}

Fig.\ref{fig:triangolo-curves} shows the split ratios over the two routes (one-hop and two-hops)
and the ratio of the total users' rate (global goodput) over the total consumed network bandwidth
(network utilisation). Let us call this ratio GCR (goodput to cost ratio).  Top plot relates to coordinated 
multi-path (CM), whilst 
bottom plot to uncoordinated (UM).
When $C_{12}=C_{31}=C$ the two problems have completely different solutions as CM
splits all traffic to the shortest-path and UM splits rates equally. At this stage GCR for CM is
30\% larger than UM's.
As  $C_{12}=C_{31}>C$ increases CM looks for more resources for the second demand 
along the two-hop path, resulting in more network cost and GCR decreases.
However using UM, GCR is insensitive to the split ratios. After a certain point CM and UM
have the same performance.

\subsubsection{Square}
\begin{figure}[htb] 
\begin{center}
$\begin{array}{cc}
\includegraphics[width=0.45\textwidth]{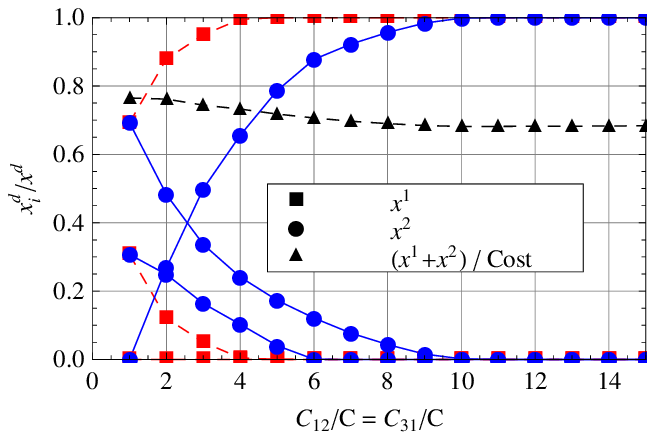}  &
\includegraphics[width=0.45\textwidth]{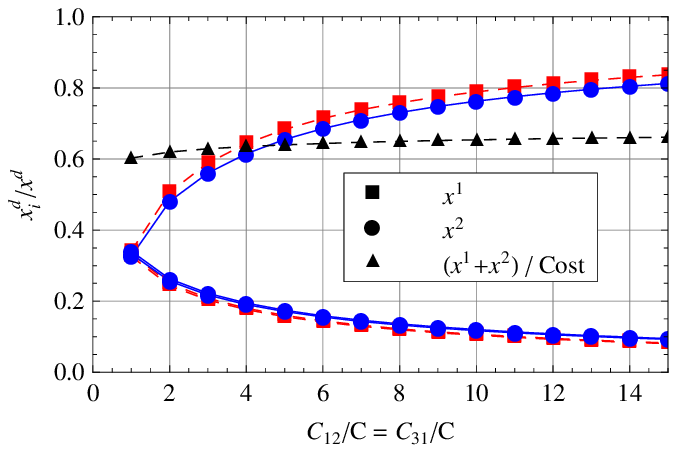} 
\end{array}$
\end{center}
\caption{Rate distribution among the available paths and user rate to network cost ratio.	}
\label{fig:quadrato-curves}
\end{figure}

In fig.\ref{fig:quadrato-curves} we have similar performance to the triangle, despite
CM needs to use more paths to attain the optimum, even to gather a small amount of
bandwidth. Furthermore GCR for CM is not that much larger than in case UM is used.
In a real protocol, secondary paths would not be used if the attained gain does not
meet the cost for the overhead that is not considered here in the model, however
significative in practice.

\subsubsection{Discussion}
MC has a larger stability region, as it consumes less bandwith to provide the same global 
goodput as UM. This might turn out not be true in practice as weakly used secondary paths
could cost to much in terms of over-head due to signalling to set up the connection,
or probes to monitor paths that are being coordinated.

\subsection{A Linear Program formulation}
\label{sec:lp}
In this section we consider  problem (\ref{eqn:opt}) in sec.\ref{subsec:optim} and select one 
particular fairness criteria: max-min. 
We write an original formulation of problem~(\ref{eqn:flow})
as an iterative linear program assuming linear costs $C(x)$. At each iteration a linear sub-problem
is solved whom, at optimum, gives the same network flow share to every demand.
This share is the maximum bandwidth that can be allocated to the most 
constrained demand. The network graph $G$ is reduced to $\tilde{G}$
through the following transformation:
$ \tilde{c}_{ij} = c_{ij} - \sum_{d\in \Gamma} x_{ij}^d$, i.e. capacities are
replaced by residual capacities after the allocation of this share of bandwidth.
if $\tilde{c}_{ij}=0$ the link is removed from the graph.
The sub-problems are formalised as follows.

\begin{align}
 \mbox{maximise} ~~~z - \sum_{i,j\in N} C\left(\frac{\sum_{d\in \Gamma} x_{ij}^d }{{c}_{ij}}\right) \nonumber
\end{align}
$\mbox{subject to}$
\begin{align}
 \sum_{k\in N} a_{ik}x_{ki}^d - \sum_{j \in N}a_{ji} x_{ij}^d = \left\{ \begin{array}{ll} z & \mbox{if}~~ i\in \mathcal{S}\\ -z & \mbox{if}~~ i\in \mathcal{E}\\ 0 & \mbox{otherwise}\end{array}\right. & \forall d\in \Gamma 
\end{align}
\begin{align}
 \sum_{d\in \Gamma} x_{ij}^d \leq \tilde{c}_{ij} & \forall i,j\in L & 
\end{align}
\begin{align}
z \leq p_d & ~~\forall d\in \Gamma ~~ \forall i\in \mathcal{S}
\end{align}

In the following section we use this iterative LP in order to obtain 
the gain that can be obtained exploiting path diversity for large networks
with a large number of demands.

\section{Analysis of large problems}\label{sec:largelp}
\subsection{Simulation set-up}
\begin{figure}[htb] 
\begin{center}
$\begin{array}{cc}
\includegraphics[width=0.60\textwidth]{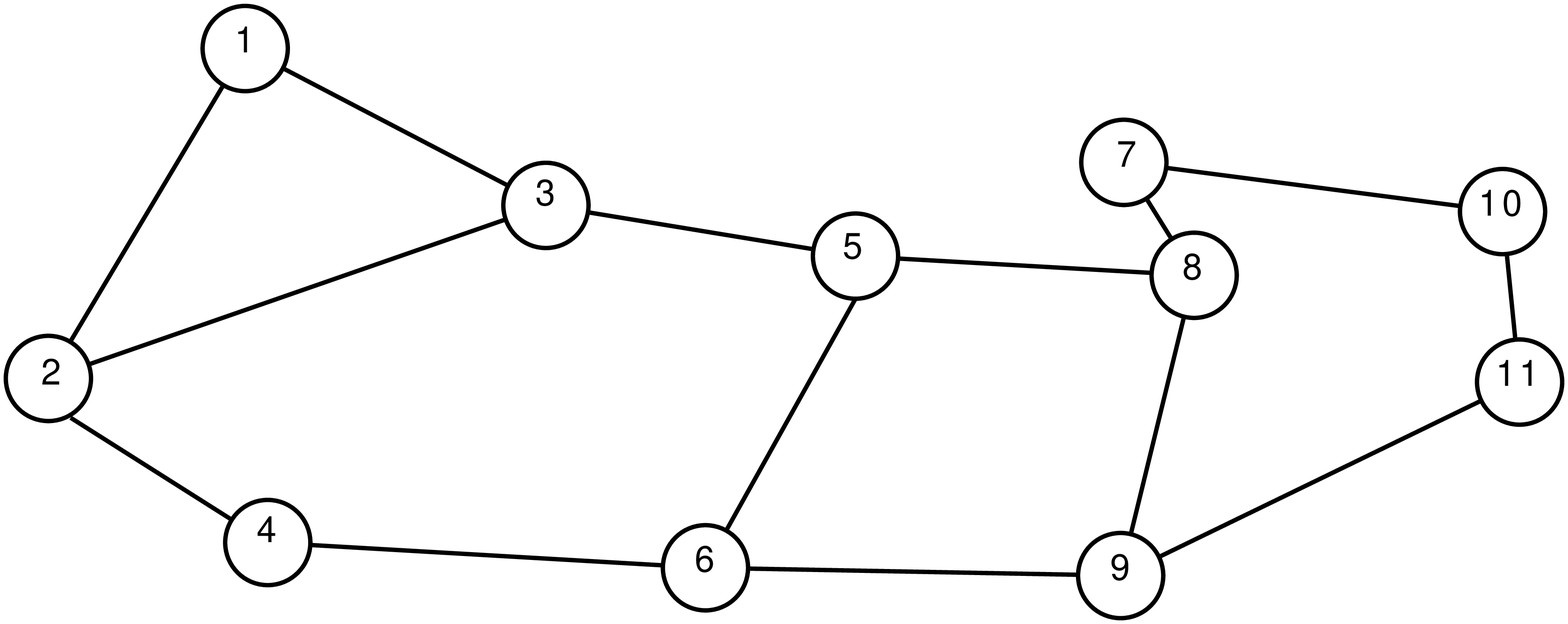} &
\includegraphics[width=0.35\textwidth,height=0.3\textwidth]{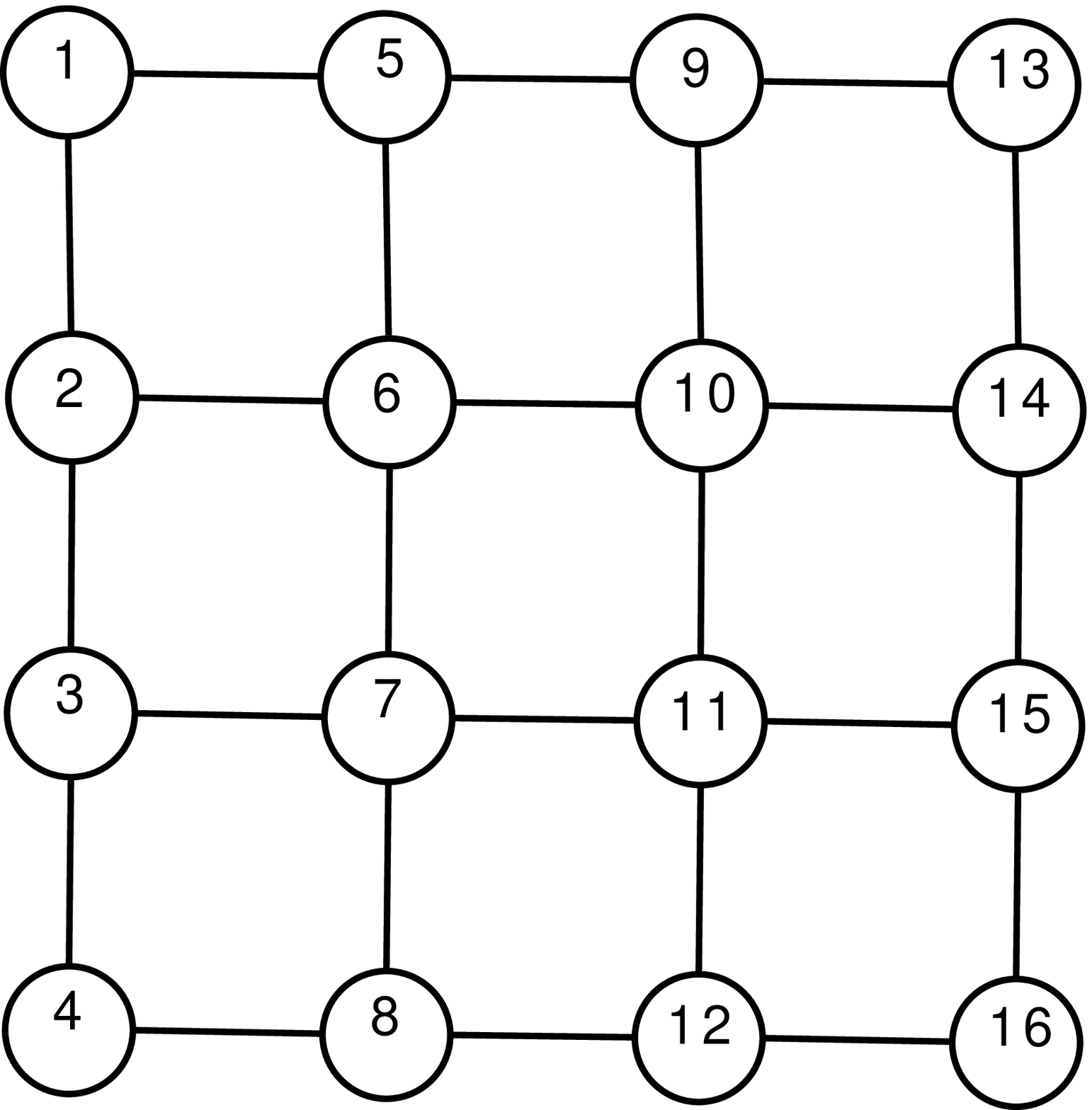}  \\
\mbox{{\bf{(a)}}} &
  \mbox{{\bf{(b)}}}\\
\end{array}$
\end{center}
\caption{(a) Abilene backbone topology; (b) a planned wireless mesh network.}
\label{fig:abilene-topology}
\end{figure}

We study two topologies as shown in fig.~\ref{fig:abilene-topology}. 
The first is the Abilene backbone network \cite{abilene}, while the second is a possible
wireless mesh network backhaul. The link capacity distribution is a Normal distribution 
with average $\bar{C}$ and standard deviation $ \bar{C}/10$.  
The Abilene topology counts $N=11$ nodes and we perform simulations with mean link 
capacity set to two scenarios: $\bar{C}=100 Mb/s$ and $ \bar{C}=50 Mb/s$ . The wireless mesh topology 
counts $N=16$ nodes and, similarly, two scenarios are considered: $\bar{C}=50 Mb/s$ $ \bar{C}=25 Mb/s$ . 
In this latter case capacities are reduced to represent radio channels with lower available bitrate.

The distribution of peak rate is taken Log-Normal with parameters $\mu=16.6$ and $\sigma=1.04$.
These values are taken from a set of fits performed on measurements gathered from Sprint backbone
\cite{nucci05} . 
Two traffic matrices are considered:
\begin{itemize}
\item {\bf Uniform}. Every node sends traffic to all other nodes. There are thus $N(N-1)$ demands where $N$ is the
number of nodes.  To match flow rates to source-destination pairs we use a technique describe in \cite{nucci05},
assuming minimum cost-path the routing used. This technique achieves the best match between a set of demands and 
a set of SD pairs connected by a given routing. This traffic pattern might be representative of an Intra-domain 
optimised backbone. Since the Abilene network has 11 nodes, in our simulations we generate a total number of 
110 demands.  We do not consider this traffic matrix over a wireless mesh topology as 
unlikely all nodes send traffic to all nodes in such networks.

\item {\bf Hot-Spot} $|S|$ nodes have $r$ flows directed to a common sink node. We fix the sink node and randomly
select $|S|$ source nodes. $|S|r$ demands are randomly assigned to these $|S|$ Source-Sink pairs.
In our experiences we set $|S|=4$ , $r=25$, for a total number of 100 demands. Node 6 is selected as sink in both 
topologies. This traffic pattern can be rapresentative of a data center located in a backbone topology or a gateway 
node in a wireless mesh network.
\end{itemize}
According to the aforementioned set-up we simulate a traffic matrix which is used as input
to the LP described in sec.\ref{sec:lp} and solved using the MATLAB optimisation toolbox.
For every scenario we evaluate the satisfaction of each demand as the ratio between its attained rate and its exogenous 
peak rate.  A demand is fully elastic if its exogenous rate is larger that the maximum attainable bandwidth
in an empty network.
Hence, satisfaction is always defined as we need not assume infinite peak rate to elastic flows.
We use this performance parameter as it is able to explicit how bandwidth is allocated with respect to the
distribution of the exogenous rates. Output data are averaged over multiple runs.\\

\subsection{Numerical results}
\begin{figure}[htb] 
\begin{center}
$\begin{array}{cc}
\includegraphics[width=0.45\textwidth]{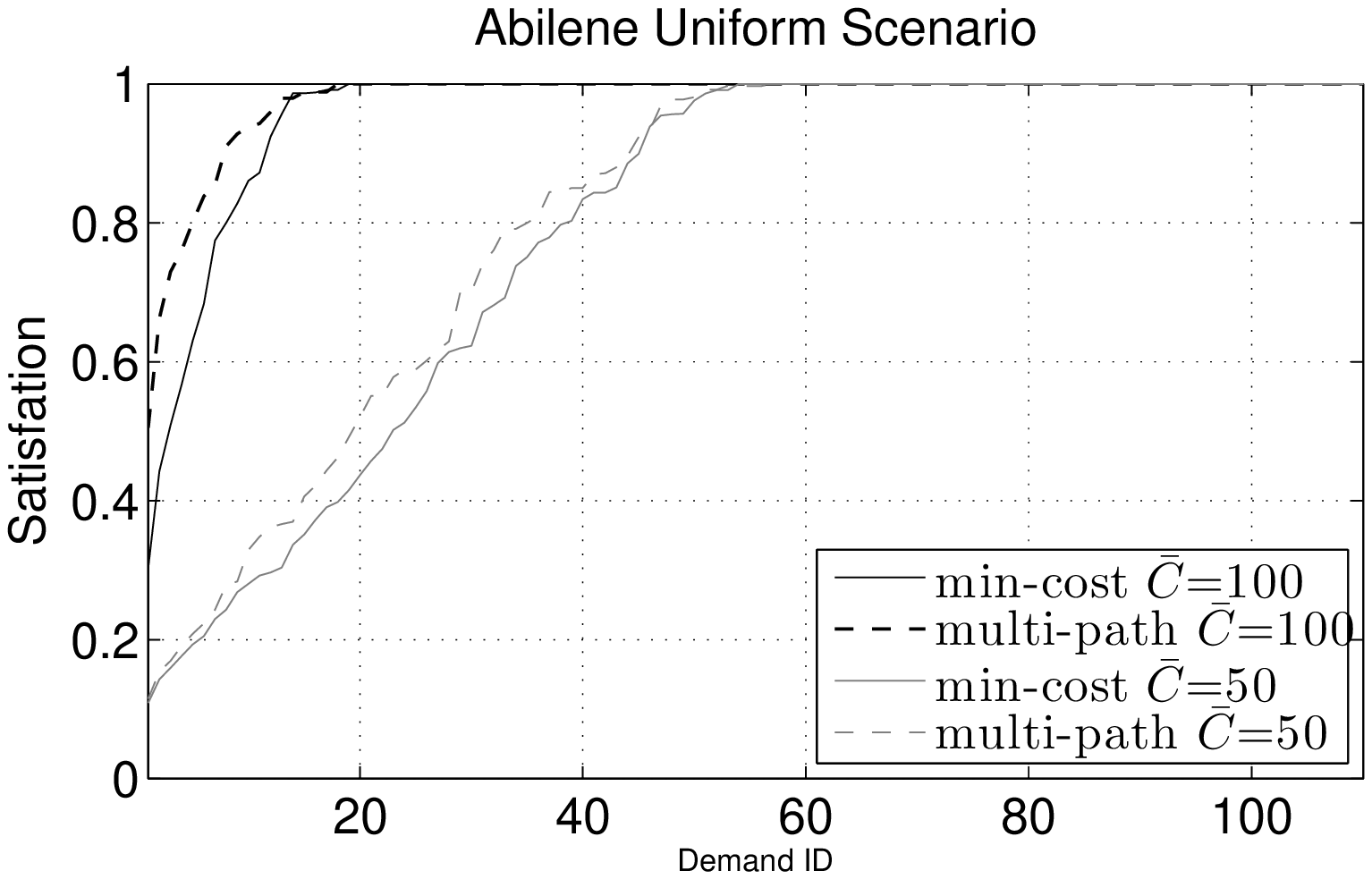} &
\includegraphics[width=0.45\textwidth, height=0.26\textwidth]{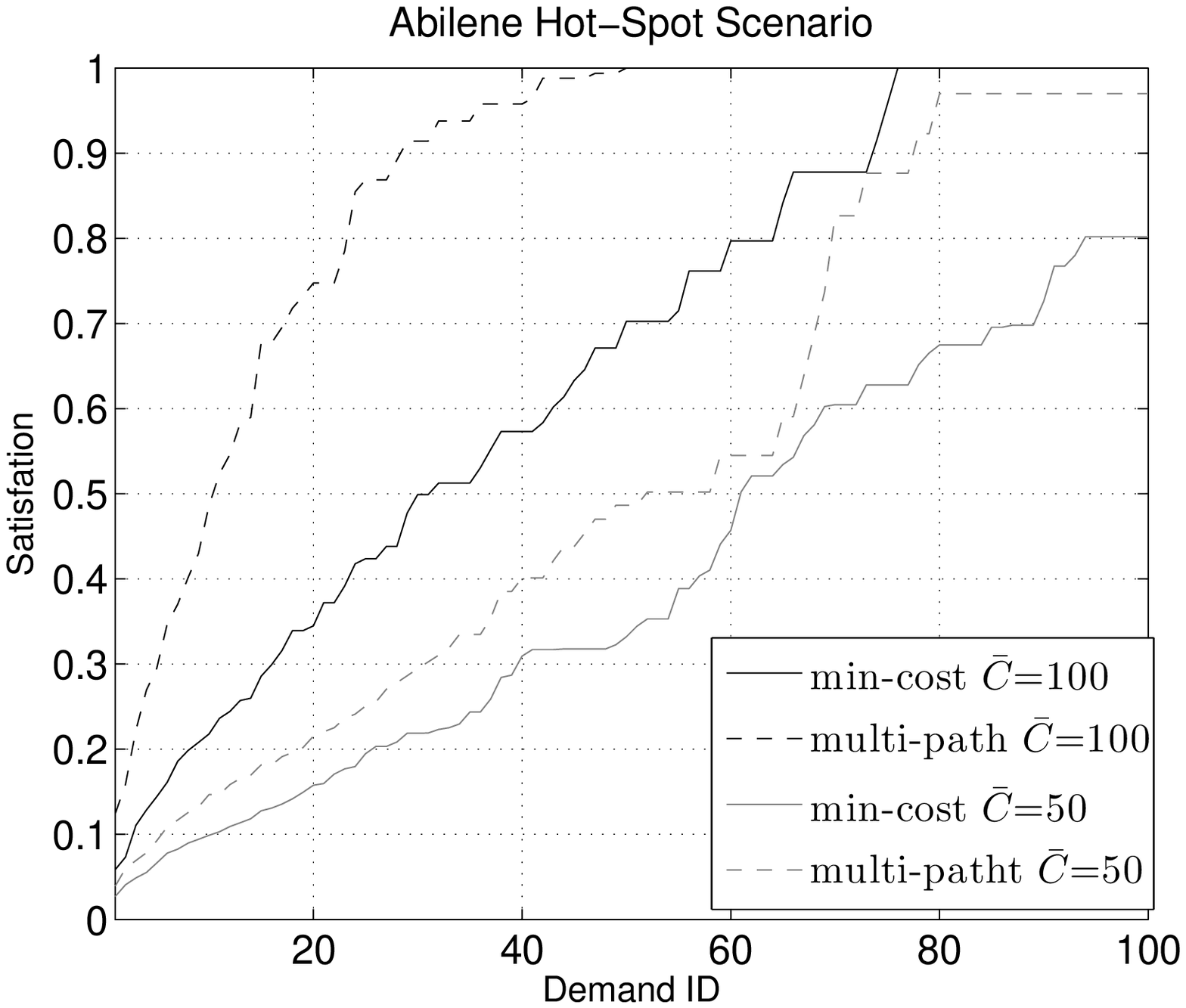} 
\end{array}$
\end{center}
\caption{Abilene topology. Satisfaction distribution for uniform (top) and hot-spot (bottom) traffic matrix.}
\label{fig:abilene-topology-curves}
\end{figure}

\begin{figure}[htb]
\begin{center}
\includegraphics[width=0.45\textwidth]{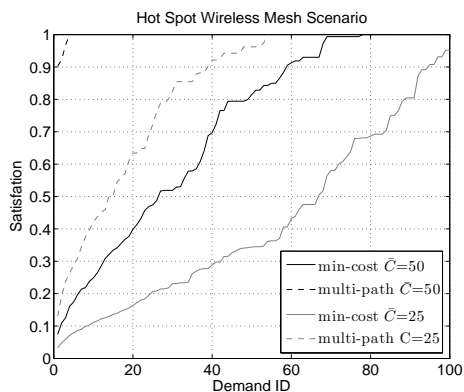} \\
\end{center}
\caption{Wireless mesh topology. Satisfaction distribution hot-spot  traffic matrix.}
 \label{fig:wireless-mesh-topology-curves}
\end{figure}

Results are reported in fig.~\ref{fig:abilene-topology-curves} and  \ref{fig:wireless-mesh-topology-curves} and,
as expected,  show multi-path routing outperforms minimum cost routing over both network topologies and 
for both traffic patterns.  However, the point is to measure the entity of the improvement.
In particular, the gain is larger for wireless mesh topology and for scenarios adopting larger link capacities.

The gain of multi-path is, in great part,  limited to flows with larger peak rate, whilst flows with lower peak rate are completely satisfied by 
both routing schemes. This because MinCost routing, under max-min fairness, penalises larger flows by fairly sharing the 
capacity of the link that acts as bottleneck among all flows there in progress.

Multi-path routing, under max-min fairness, acts similarly except that flows can retrieve bandwidth, not only on their minimum cost 
path, but also on their secondary routes. Indeed, simulations show that low rate flows do not take any advantage of path diversity,
while high rate flows retrieve additional bandwidth along other paths. 

Max-min multi-path routing avoids to use more than one path to 
route low rate flows. This has beneficial effect in practice as it avoids wastage of resources due to the overhead,
that might be justified only above a certain minimum rate.

Fig.~\ref{fig:abilene-topology-curves} reports results for the Abilene network. The gain of max-min multi-path routing is 
very large for hot-spot traffic matrices, and even more significant when network capacities are larger. The routing scheme
is able to transfer 65\% additional traffic when $\bar{C}=100$ and 41\% when $\bar{C}=50$ , with respect to 
minimum cost routing. However, under uniform traffic, the gain is much smaller; i.e. only 4\% when $\bar{C}=100$ and 3\% when $\bar{C}=50$.

Under uniform traffic, demands are spread all around nodes and even high rate flows do not exploit path diversity as they are almost 
completely satisfied over their minimum cost path. 

If resources become scarce high rate flows cannot exploit path diversity because all links are already saturated by flows routed along 
their minimum cost paths. Therefore as the capacity is increased, the gain is distributed to low rate flows first, and
to those with higher rate at last.

Of course, as the capacity is fairly large, resulting in a lightly loaded network, both schemes 
have similar performance as all flows are satisfied along the mincost route.	

The variance of the satisfaction is quite small, and between 0.05 and 2.3 for all scenarios. 
However it is not uniformly distributed among all flows. 
In fact, large flows are affected by larger variance than small flows. This is due to the 
max-min fairness criteria that allocates a minimum amount of bandwidth to all flows. This always assures complete 
satisfaction for small flows while satisfaction of large flows depends on the network capacities that varies from simulation to 
simulation.

\section{Architectures and protocols}
\label{sec:arcprot}
\subsection{Architectures}
\label{sec:archs}
Any of the proposals for congestion control, and congestion control exploiting path diversity, can be restated in the
context of an optimisation problem of the kind of (\ref{eqn:opt}) or the uncoordinated counterpart (\ref{eqn:opt2}).
Such problems lead to different architectures based on different theoretical foundations.

We divide such architectures in three groups: fully decentralised, quasi decentralised, flow aware.
This classification is for the ease of exposition and for sake of clarity, however it lays itself open to critics.

\begin{itemize}
\item{\bf Fully decentralised (FD)}. 
In such architecture network nodes store, switch and forward packets from an input to an output interface.
Neither scheduling nor active queue managament is implemented into the nodes.
A form of cooperation is assumed between sources, that implement a common rate control algorithm and, 
if needed, a common multiple path splitter. In general sources need to be conformant to a common fairness
criteria. This  is the case for the current Internet and multi-path TCP \cite{Srikant06} is one example of such
controller.

\item {\bf Quasi decentralised (QED)}. 
Nodes implement any form of active queue management (AQM),  to prevent congestion, and explicit
congestion notification (ECN) that is triggered according to some temporisation. ECN might also be result of a local node
calculation. Sources exploit such notification to control sending rate and split decisions.
\cite{Rexford07, RexfordCo07,katabi05} are two examples in the context of intra-domain traffic engineering.

\item {\bf Flow aware (FA)}. Nodes implement packet schedulers that realise fair bandwidth sharing between flows. Sources
balance traffic among the available routes subject to the restriction imposed by packet schedulers and autonomously
decide how to exploit as better as they can network resources. Congestion control and bandwidth
allocation are solved separately. \cite{oueslati07} propose a routing scheme inspired by the the technique
of "trunk reservation" used in PSTN to allocate circuits to secondary paths in case the direct paths were experiencing
overload.  \cite{oueslati07} 
\end{itemize}

Each of above-mentioned architectures have its counterpart that make no use of path diversity. 
It appears difficult to unequivocally select one approach. For instance, the question of the deployment
of a certain congestion control algorithm is still hotly debated (see the newsletter \cite{Touch}).

The use of experimental protocols (e.g. cubic in Linux) scares, as {\bf{FD}} architectures assume a 
common form of cooperation between users and anarchy might be costly, 
whether degenerates in congestion collapse, or being less negative, in unfairness in sharing resources. 

Manifest truth is that {\bf{FD}} are very simple and do not require any sort of parametrisation, especially at nodes,
and extensions that make use of path diversity are easy to deploy as source routing is made available. However,
the main concern  that obstructs the deployment of source routing is security.

{\bf QD } enhances {\bf{FD}} and make it more efficient and stable. XCP \cite{katabi02}  is an example of protocols
of this kind while, in the context of intra-domain TE, TEXCP \cite{katabi05} and TRUMP \cite{Rexford07} 
belong to this class.

\subsection{Flow aware architecture}
\subsubsection{Per-flow fair queueing}
The benefit of per-flow fair queueing has long been recognised \cite{shenker89, nagle85}. 
Besides, it is robust againts unfair use of resources because of non standard conformant use
of network transport protocols resulting from bad implementations (rare), 
absence of common agreement (the case of cubic in Linux \cite{Touch})
as well as malicious use that exploits others' weakness (more aggressive
congestion controllers for instance).
Per-flow fair queueing relieves the network to assume standard conformance of end to end protocols.
Assured fairness allows new transport protocols to be introduced without relying on detailed
fairness properties of preexisting algorithms.

\subsubsection{Overload control}
Per-flow fair queueing is feasible and scalable in presence of overload control
\cite{korte04, korte05}. When demand exceeds capacity the scheduler assures
equal performance degradation to all flows. 
However, arising congestion at flow level is a transient phenomena, as  users would 
quit the service in crowds bringing back utilisation to normal loads. This is perceived 
by the user as service break-down.

At present, no overload control is implemented, in any form, within the network.
At a certain extent, this is assured within the ISPs backbone, and in part within the access, by
over-provisioning link capacities according to an estimated traffic matrix 
(using netflow or similar tools). Such methodology does not solve local congestion
in small periods  of time and force users to quit profiting from a service.

In wireless mesh backhaul overprovisioning would not be anymore a feasible solution
and multi-path routing might be necessary.

Overload control has to be dynamic and fast reacting to congestion. This can be realised
looking for resources from other available ways, for instance the availability  multiple routes
to join a service.

\subsubsection{Per-flow path selection and admission control}
A sub optimal approach is to select one single path among many others.
In presence of a number of paths to reach a destination, a flow can be deflected to a better route,
e.g. with larger fair rate or with minimum cost. Such a greedy scheme is not stable 
and would led to oscillations if the driving metric is not stable enough. This is the case
of fair rate.

\subsubsection{Per-flow multiple path routing}
The presence of per-flow fair queueing in every link impose per-path fair bandwidth sharing.
This means that the utility of a single user is not a function of the total attained rate as 
he is not allowed to get more bandwith of the fair share along its minimum cost path.
Therefore the utility is given by the sum of the utility of each singular path as (\ref{eqn:opt2}).
We know that in general this problem is suboptimal with respect to a coordinated splitter.
In this case the optimum attained rate is given by $x_i^d = \partial_{x_i^d} U(q_i^d)^{-1}$
being $q_i^d$ the cost of path $i$ for user $d$. The split ratio is the inversely proportional to the
cost. 
In a recent paper Key et al. \cite{Key06} prove that in absence of coordination,  the stability region
of the number of flow in progress in the network, is reduced. This is shown for a triangle topology
and uniform traffic matrix. The flow model assumes that flows arrive according
to a stochastic process and leave the system after being served an amount of data which is distributed.
A reduced rate region at flow level is a consequence of  (\ref{eqn:opt2}) as, the same rate
is obtained at a larger cost.  As we show in the example \ref{sec:triangolo}.
Notice also that in presence of uniform traffic matrices, the over-head requested by multi-path is not really
justified with respect to minimum cost path selection as there is almost no gain, as we have shown in
the previous section.

\subsection{Multi-path Iterative Routing Traffic Optimizer}
In this section we propose one of the main outcome of the paper: Multi-path Iterative Routing Traffic Optimiser (MIRTO), 
a fully distributed algorithm aimed at obtaining, 
in a decentralised way, a multi-path optimal strategy proposed in sec.\ref{sec:lp}.

It is designed around TCP (AIMD) and it allows coordinated traffic split over multiple paths. 
MIRTO might likely run on end-hosts and can work in any of the architectures described in sec.\ref{sec:archs}.

In order to run MIRTO, hosts should discover available routes and link capacities along them to reach a destination.
The way such information is collected is out of the scope of our algorithm but they can for example be discovered 
through link-state routing protocols. The algorithm works even whether information is incomplete but  in a less effective way,
as in case paths are partially discovered and path diversity limited.

Flow splitting over multiple paths can be performed in several ways according to the application context. 
By means of path computation capabilities of MPLS for intra-domain TE, or through middlewares running
on proxy nodes active as middle-layer. 

\begin{table}[ht]
    \centering
        \begin{tabular}{|l|l|}
        \hline
        \textbf{Symbol} & \textbf{Meaning}\\
        \hline
        $\Delta^+$ & positive step $\in \mathbb{R^+}$\\
        \hline
        $\Delta^-$ & negative step $\in \mathbb{R^-}$\\
        \hline
        $Q^d_k(t)$ & cost of $k^{th}$ path of flow $d$ at time $t$\\
        \hline
        $RTT_k^d$ & Round Trip Time of the $k^{th}$ path of flow $d$\\
        \hline
        \end{tabular}
\caption{Algorithm's notation}
\label{tab:notation_algo}
\end{table}

Algorithm~\ref{algo:MIRTO} shows MIRTO pseudo-code for rate control over a given path $k$ for a given flow $d$.  
Notation is reported in table \ref{tab:notation_algo}.
Operations are performed every $RTT_k^d$  or in general when new information are available on the state of  path $k$. 
Path costs are computed according to (\ref{eqn:MIRTO_cost}). They only depend on capacity of the link along the route.
Indeed, in case of linear cost (\ref{eqn:cost2}) is equal to $\sum_{ij \in L^d_k}{\frac{1}{C_{ij}}}$.
Link with larger capacity are better ranked regardeless of utilisation.
Infinite cost in (\ref{eqn:MIRTO_cost}) just indicates congestion notification \`{a} la TCP and the  route is marked congested.
\begin{align}
Q^d_k(t) = \left\{ \begin{array}{ll} \sum_{ij \in L^d_k}{\frac{\Delta^+}{C_{ij}}}  & \mbox{if}~~\forall (i,j)  \in L_k^d~~\rho_{ij}(t)<C_{ij} \\
 \infty & \mbox{if}~~\exists (i,j) \in L_k^d ~~\rho_{ij}(t) \ge C_{ij}\\ \end{array}\right.
\label{eqn:MIRTO_cost}
\end{align}

\begin{algorithm}
\begin{algorithmic}
\FOR{$k\in P^d$}
\STATE compute $Q^d_k(t+RTT_k^d)$
\ENDFOR
\IF{$Q^d_k(t+RTT_k^d)=\infty~~\forall k \in P^d$}
\FOR{$k\in~P^d$}
\STATE $x_k^d(t+RTT_k^d) \leftarrow x_k^d(t)-x^d(t)\Delta^-$
\ENDFOR
\ELSIF{$p^d(t) \ge x^d(t)$  \textbf{and} $Q_k^d(t+RTT_k^d)=\min_{k}{Q_k^d}(t+RTT_k^d)$}
\STATE $x_k^d(t+RTT_k^d)  \leftarrow x_k^d(t)+\Delta^+$
\ELSIF{$p^d(t) < x^d(t)$  \textbf{and} $Q_k^d(t+RTT_k^d)=\max_{k}{Q_k^d}(t+RTT_k^d) $}
\STATE $x_k^d(t+RTT_k^d)  \leftarrow x_k^d(t)-\Delta^-$
\ENDIF
\end{algorithmic}
\caption{\label{algo:MIRTO}MIRTO algorithm for a given demand $d$}
\end{algorithm}

An increase of $\Delta^+$ is done over the minimum cost path when there is at least one 
non congested path,  and the total flow rate can still get increased if lower than its own peak rate. 
The rate increase is chosen constant and independent to flow rate not to favour higher rate flows.
This increase can be considered as a probe in order to discover the global optimum split.
In principle all path might be probed as, when those better ranked are congested, worse ranked 
routes can be exploited. The importance of probing in such kind of controller to avoid
to be trapped in non optima equilibria has been highlighted in \cite{Kelly05}.

Flow rate is decreased in two ways according to network conditions. Firstly MIRTO reduces 
sending rate over all paths  when they are all congested. The rate decrease is proportional to 
the total flow rate. This guarantees fairness. Otherwise, small flows or new coming flows starting with lower rates 
would be disadvantaged with respect to higher flows. Moreover decrement is performed over all paths 
at the same time  in order to allow flow split re-arranging. 

In fact, after this decrease, there is newly available bandwidth and flows will grow according to the
aforementioned rules. So,  if the current split is only locally optimum the controller would move towards
a global optimum according to this search strategy.
On the other hand, a rate decrease is performed any time the total flow rate is larger than flow peak rate 
and at least one non-congested path is available. This means that,  as a better ranked route
is willing to increase its rate, this must be done to the detriment of a worse ranked route,  without
altering the global rate which is bounded by the exogenous rate.
In fact even if the total flow rate is equal to peak rate, the path splitting could be only locally optimum
i.e. more expensive.  It worth recalling that a more expensive path might also be characterized 
by larger end to end delays. If more than one maximum/minimum cost paths exist, the decrease/increase 
is shared among them.

This mechanism allows MIRTO to reach a global optimum in presence of demands with or without 
exogenous rate as it follows the classical water filling procedure that lays at the base of the max-min 
fairness criteria.

The selection of the minimum-maximum cost path when performing increase/decrease operations 
guarantees coordination between different paths of the same flow as long as the network does not 
impose any additional fairness semantic.

\subsection{Stability and optimality}
MIRTO  can be described through a fluid equation that approximates its behaviour and can be used
to prove convergence and stability. We suppose first no network delay and then we generealise to
the realistic case.
\subsubsection{Absence of network delays}
First consider the case the flow has no exogenous rate.
$$ \dfrac{d x_{i}(t)}{dt} = \Delta^+[1-q(t)]\kappa_i(t)-\Delta^-q(t) \sum_{k=1}^{N}x_{k}(t) $$
$\kappa_i(t)$ is the probability that $i$ is the minimum cost  path at time $t$. 
$q(t)$ is the probability that all path are congested. Therefore at steady state
$$\sum_{k}x_{k}(\infty) =\kappa_{\max}(\infty)\frac{1-q(\infty)}{q(\infty)}\frac{\Delta^+}{\Delta^-}$$
 where 
$\kappa_{max}(\infty) = \max_k \kappa_j(\infty) $. This means  traffic is  split
among minimum cost paths, possibly a single path. All the others
are subject to continuous probing such that $x_i(t) \sim \kappa_i(t)$. 
This feature allows the controller not to be trapped in equilibria points
that are not optima. A path is probed as frequent as it is ranked the best
among the others. 
This can be seen from simulations in Section~\ref{sec:case} in fig. \ref{fig:mirtoFD}.
In case the flow is peak rate limited the previous equation can be rewritten as follows.
\begin{eqnarray}
\dfrac{d x_{i}^d(t)}{dt}= \{\Delta^+\kappa_i(t)s^d(t)- \Delta^-r_i(t)[1-s^d(t)]\}[1-q^d(t)]+-\Delta^-q^d(t) \sum_{k}x_{k}(t)\nonumber 
\end{eqnarray}
where $s^d(t) =Pr[x^d(t)>p^d]$ and $r_i(t)$ is the probability that $i$ is the more expensive path.
Convergence can be discussed as in the previous case when there were no peak rate.
\subsubsection{Presence of network delays}
In presence of network delays  $\kappa(t)$, $q(t)$, $r(t)$ and $s(t)$ are delayed information 
at the source. A problem of stability in the sense of theory of control arises. We do not 
provide here rules on how to set  $\Delta^+$ and $\Delta^-$ in order to keep the system asintotically
stable in presence of delays. Using standard techniques as described in \cite{Srikant04}
this can be easily obtained for simple topologies. Cumbersome calculations, and
the use of the generalised Nyquist criterion can be used to prove stability for a general topology.

\subsection{A case study}\label{sec:case}
In this section we evaluate the performance of the above mentioned architectures by means 
of fluid simulations in order to diplay the convergence behaviour of three selected scenarios. 
Each scenario represents one of the three architectures  considered in Section~\ref{sec:archs}

\begin{figure}[htb]
\begin{center}
$\begin{array}{cccc}
\includegraphics[width=0.2\textwidth,height=0.2\textwidth]{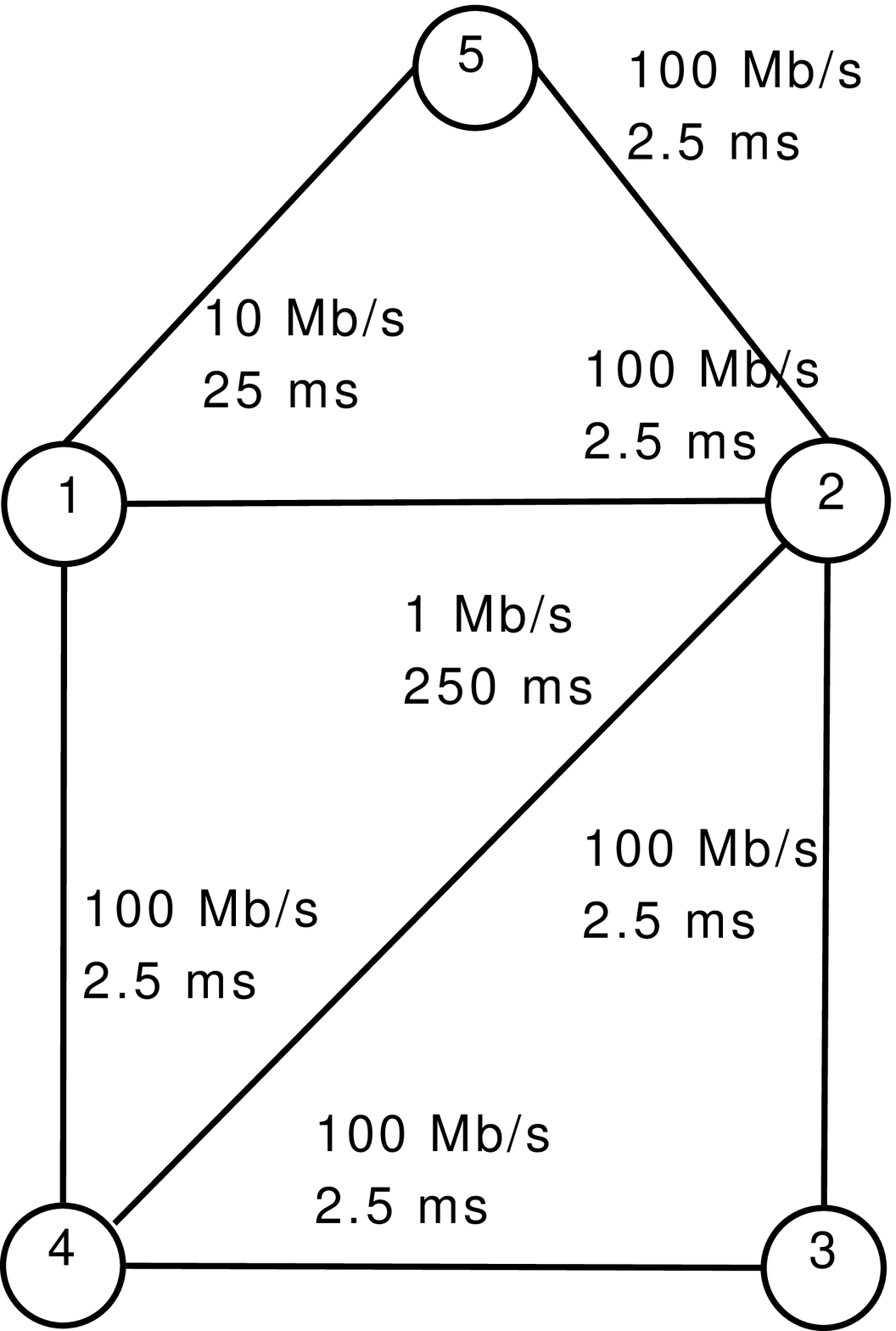} &
\includegraphics[width=0.2\textwidth,height=0.2\textwidth]{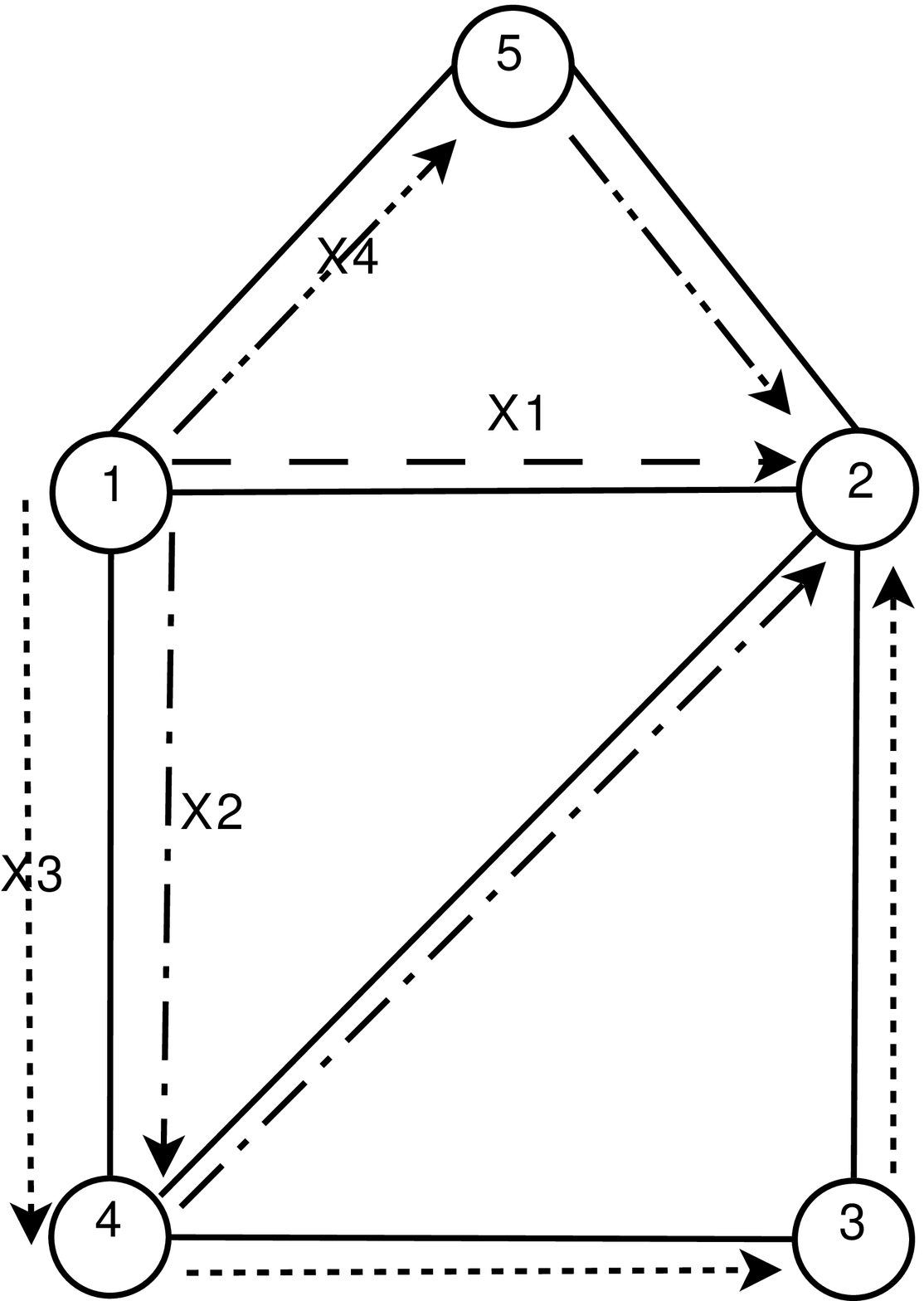}&
\includegraphics[width=0.2\textwidth,height=0.2\textwidth]{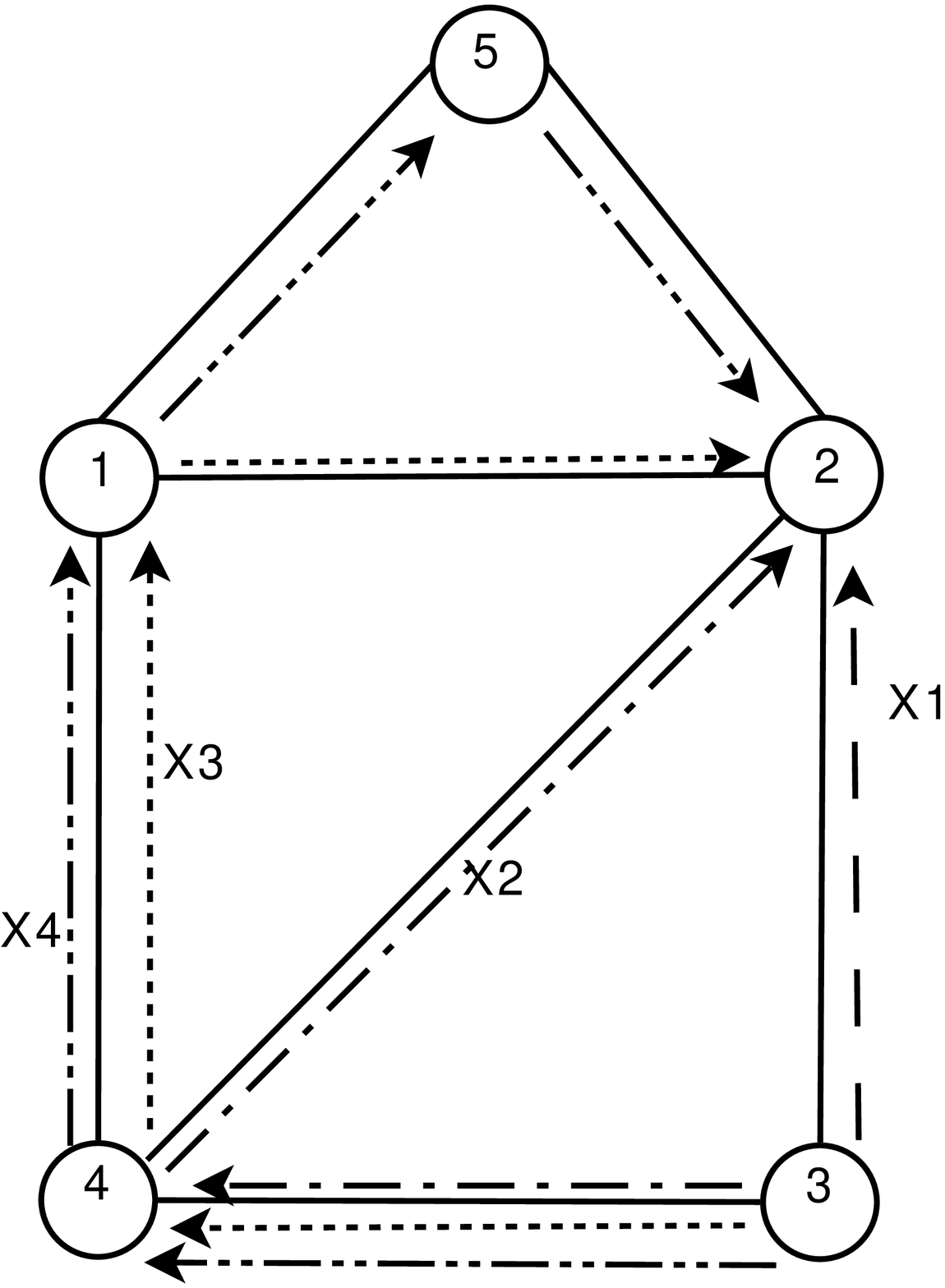} &
\includegraphics[width=0.2\textwidth,height=0.2\textwidth]{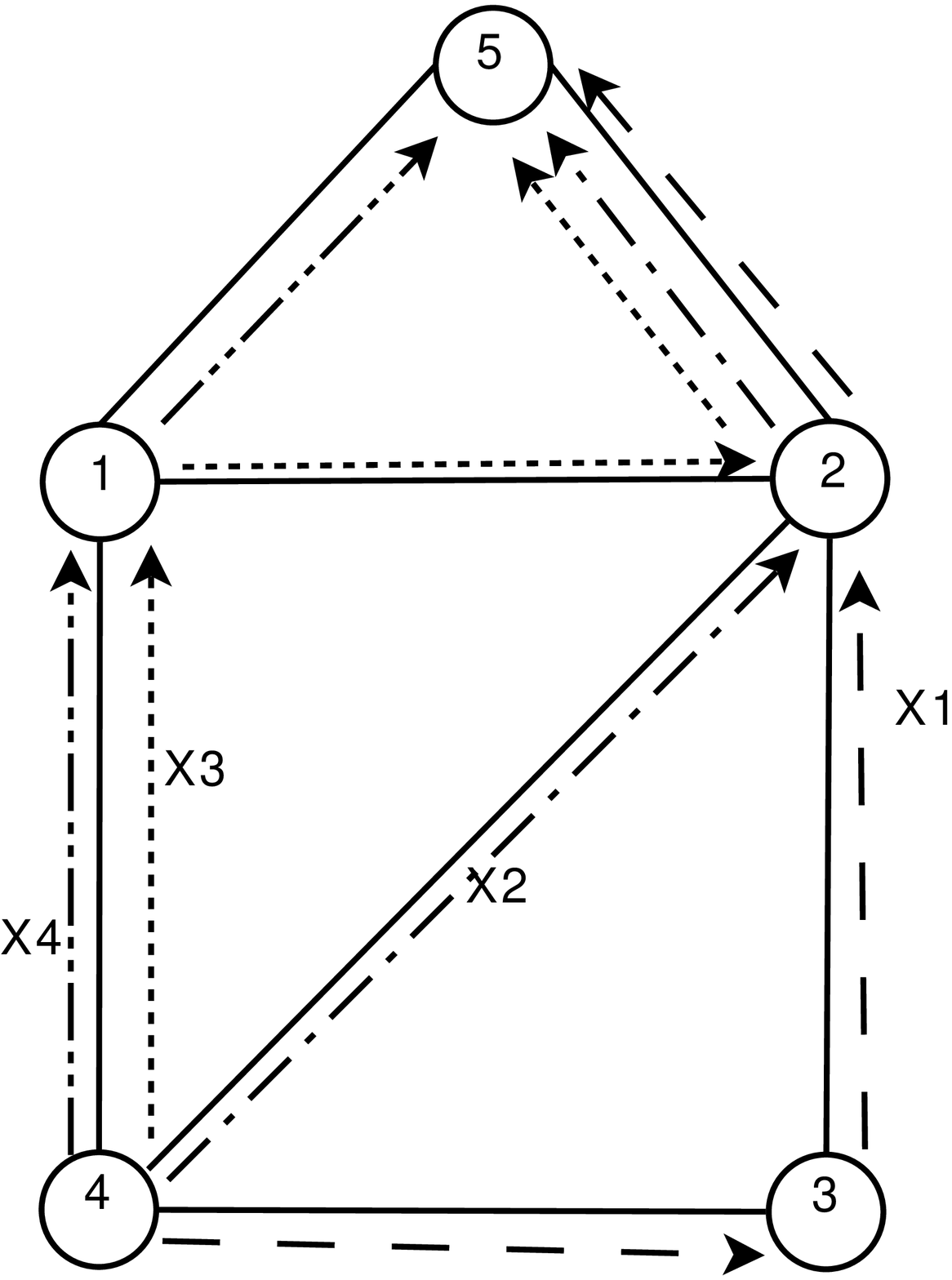}\\
\end{array}$
\end{center}
\caption{Simple network topology with the set of possible path for each of the three demands.}
 \label{fig:simple-topology-access}
\end{figure}

\subsubsection{Simulation setup}
The FD and the FA architectures are analyzed by supposing nodes run the MIRTO algorithm. Recall MIRTO has been specifically 
designed for FD solutions and requires to set rate increase and decrease values. They should allow the algorithm to overcome local 
optimal split to attain the global optimum and, at the same time, limit traffic fluctuation.  The problem of setting increase-decrease 
values is a well known trade-off of TCP and TCP-like protocols and MIRTO inherits it as well. In our simulations we set them to 
$\Delta^+=0.5 Mb/s$ and $\Delta^-=0.013$.\\

As concern QD architectures, we implement a modified version of the TRUMP \cite{RexfordCo07} algorithm. It differs from the original 
one simply because it can work even in presence of flows with a given peak rate. This is achieved by decreasing the rate of a flow over 
all paths when its total rate overcomes its peak rate. TRUMP requires to set three parameters.  A first parameter, called $w$ is a 
weight to adjust balance between utility and cost function. It is strongly related to topology and link capacities and it tunes the maximum 
network load. A second one, called $\beta$, weighs the impact of a congested link, while a third one, called $\gamma$, is the 
increase/decrease step. For our comparison purposes we set them to $w=10^{-2}$ to allow flows to fully utilize links, $\beta=10^{-3}$ 
to respect link capacities and $\gamma=10^{-3}$ to limit rate oscillations.

\begin{table}[ht]
    \centering
        \begin{tabular}{|c|c|c|c|c|c|c|c|}
        \hline
         & \multicolumn{7}{|c|}{\textbf{Time [sec]}} \\
        \hline
         & 0-5 & 5-10 & 10-18 & 18-25 & 25-40 &  40-60 & 60-80\\
        \hline
        $X^1$  &  0 & 0 & 70 & $\infty$ & $\infty$ &  $\infty$ & $\infty$\\
        \hline
        $X^2$ &  0 & 30 & 30 & $\infty$ & $\infty$ &  $\infty$ & $\infty$\\
        \hline
        $X^3$ & 50 & 50 & 50 & 50 & 50 &  $\infty$ & 55\\
        \hline
        \end{tabular}
\caption{\label{tab:rate_evolution}Per user peak rate evolution over time. Rates values are expressed in Mb/s.}
\end{table}

\subsubsection{Results}
Figure \ref{fig:simple-topology-access} reports the topology used in the simulations we show in this section to
show the behaviour of MIRTO. Link capacities and latencies have been selected in order to have path diversity 
and different response time.  
There are three flows in the network and we select as source-destination pairs 1-2, 3-2 and 4-5. 
Flow peak rates change over time as specified in table ~\ref{tab:rate_evolution}. 
Note that, $\infty$ is used to indicate elastic flows. 
This just means flow peak rate is larger than the available network resources. 
Every flow can be split over four different paths as shown in Figure. \\

\begin{figure}[htb]
\begin{center}
$\begin{array}{cc}
	\includegraphics[width=0.5\textwidth,height=0.3\textwidth]{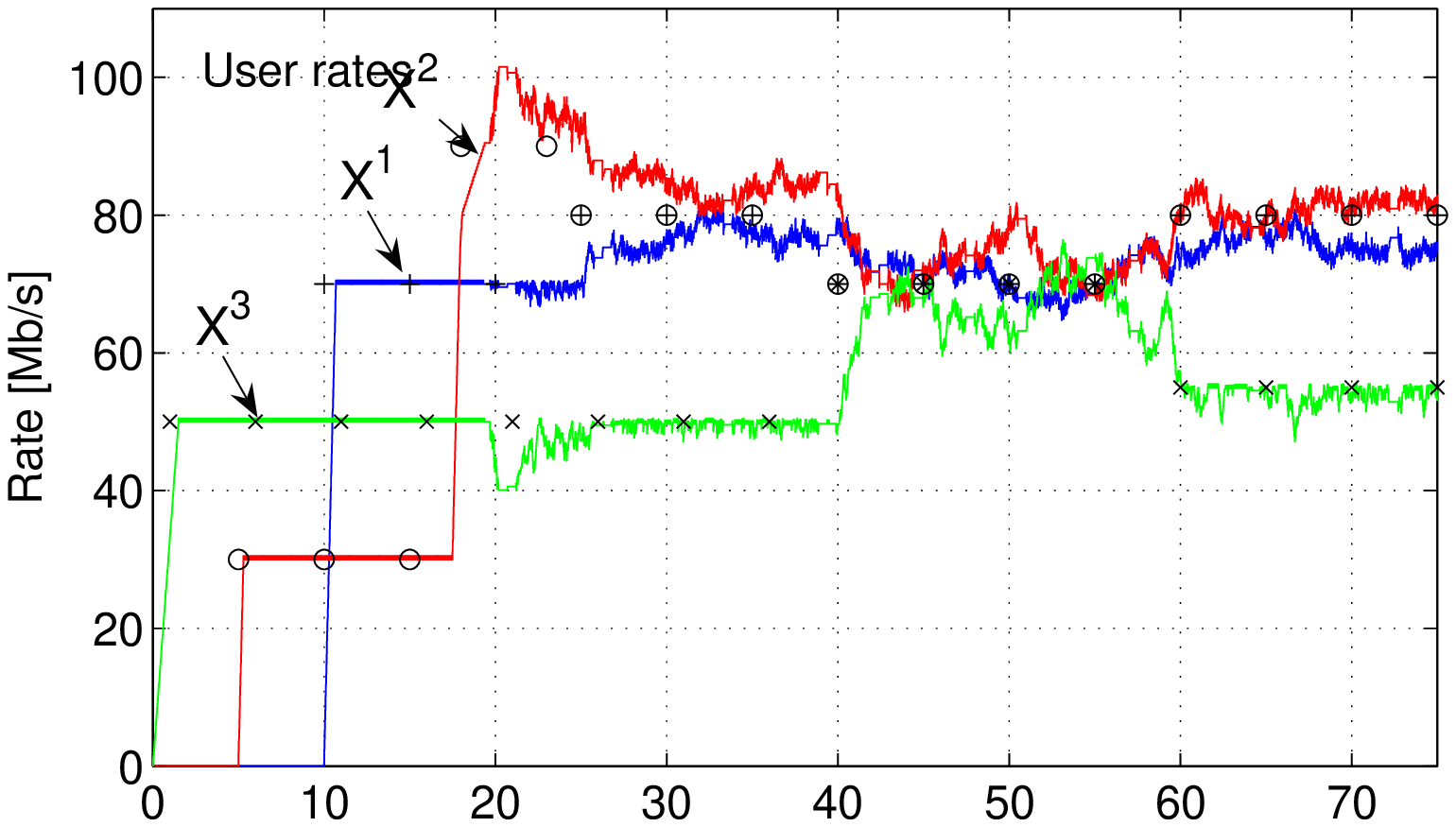} &
\includegraphics[width=0.5\textwidth,height=0.3\textwidth]{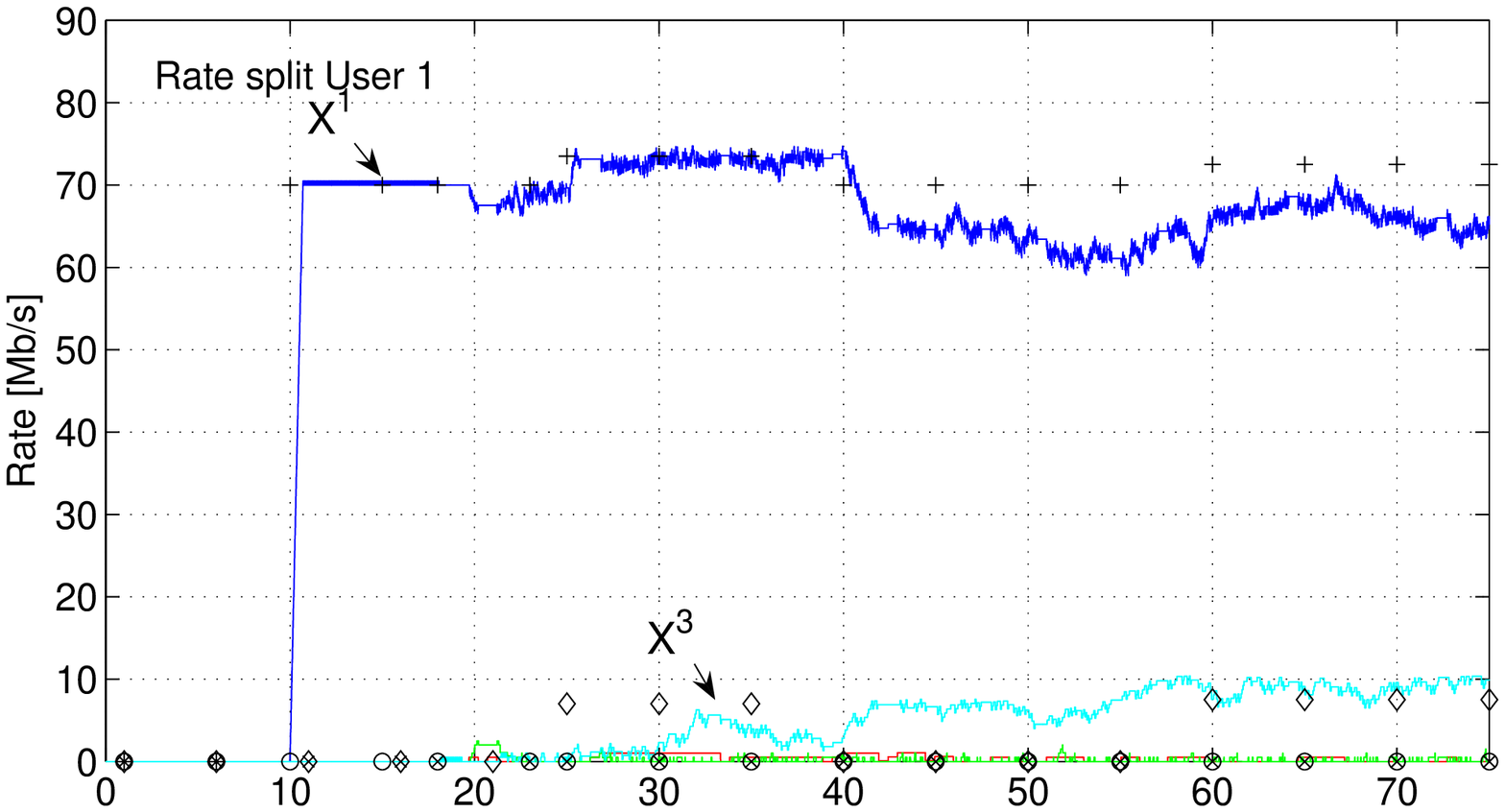}\\
\includegraphics[width=0.5\textwidth,height=0.3\textwidth]{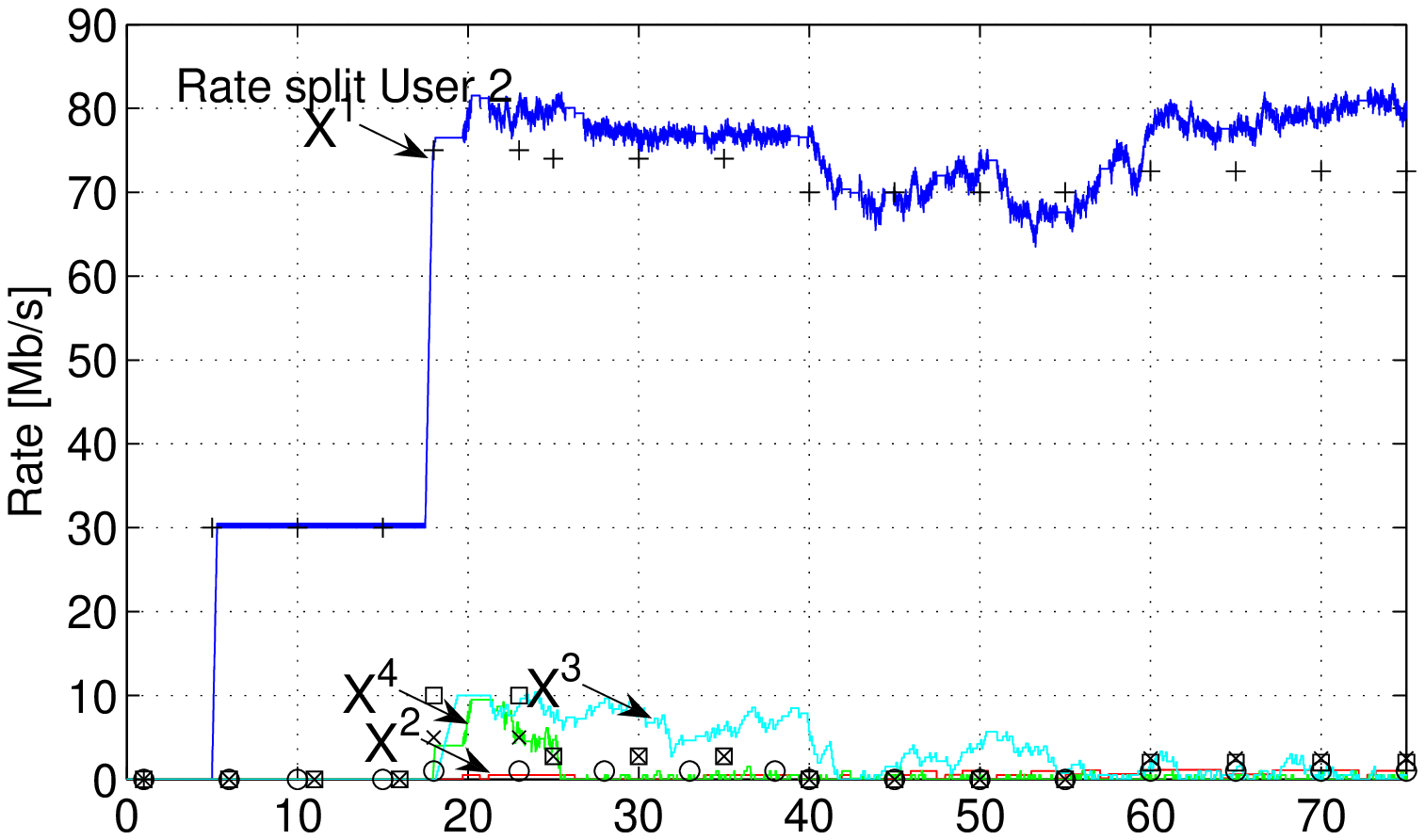} &
\includegraphics[width=0.5\textwidth,height=0.3\textwidth]{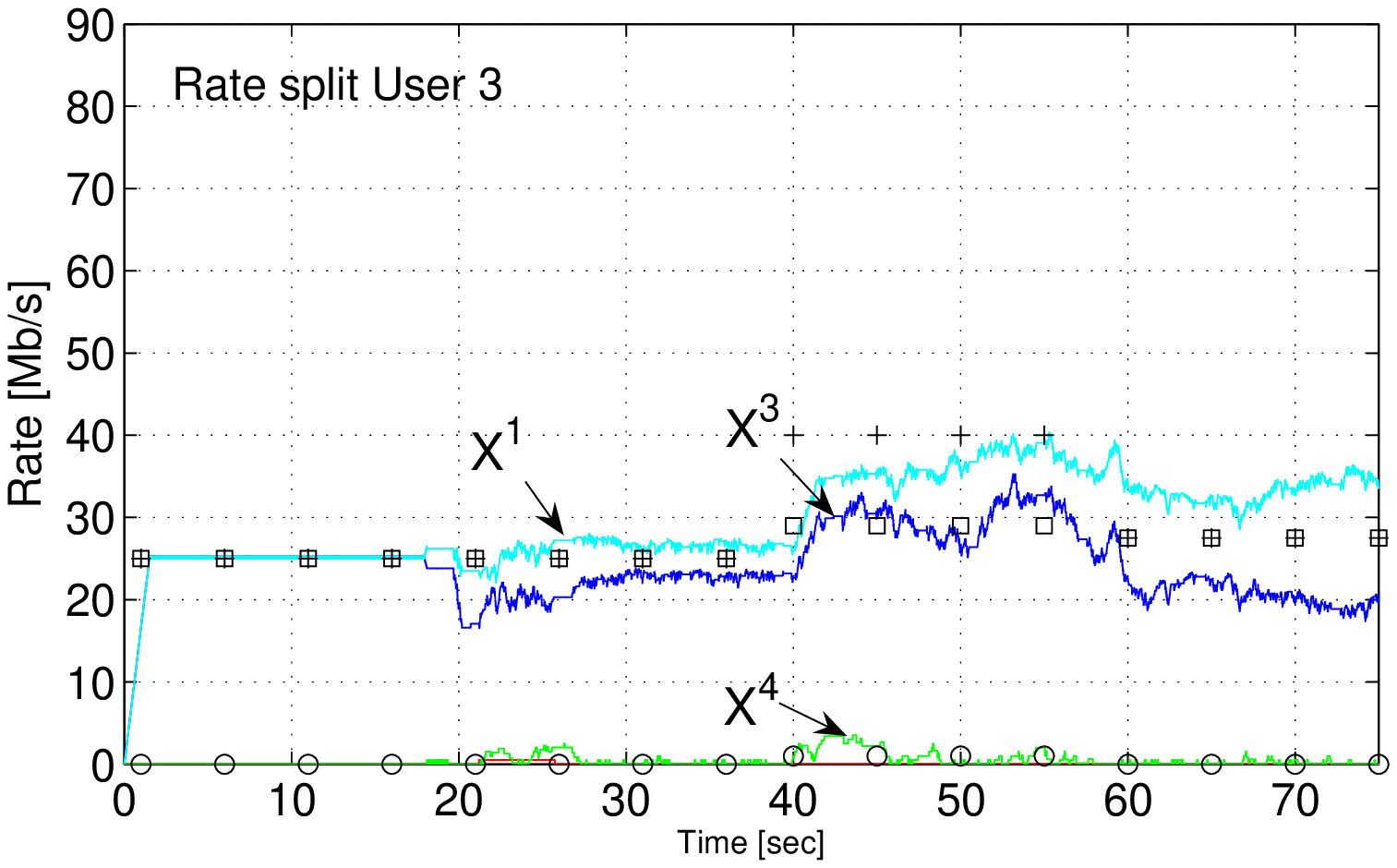}\\
\end{array}$
\end{center}
\caption{Time evolution of the rate allocation for MIRTO in a FD architecture. Points indicate the optimal allocation.}
 \label{fig:mirtoFD}
\end{figure}

\begin{figure}[htb]
\begin{center}
$\begin{array}{cc}
\includegraphics[width=0.5\textwidth,height=0.3\textwidth]{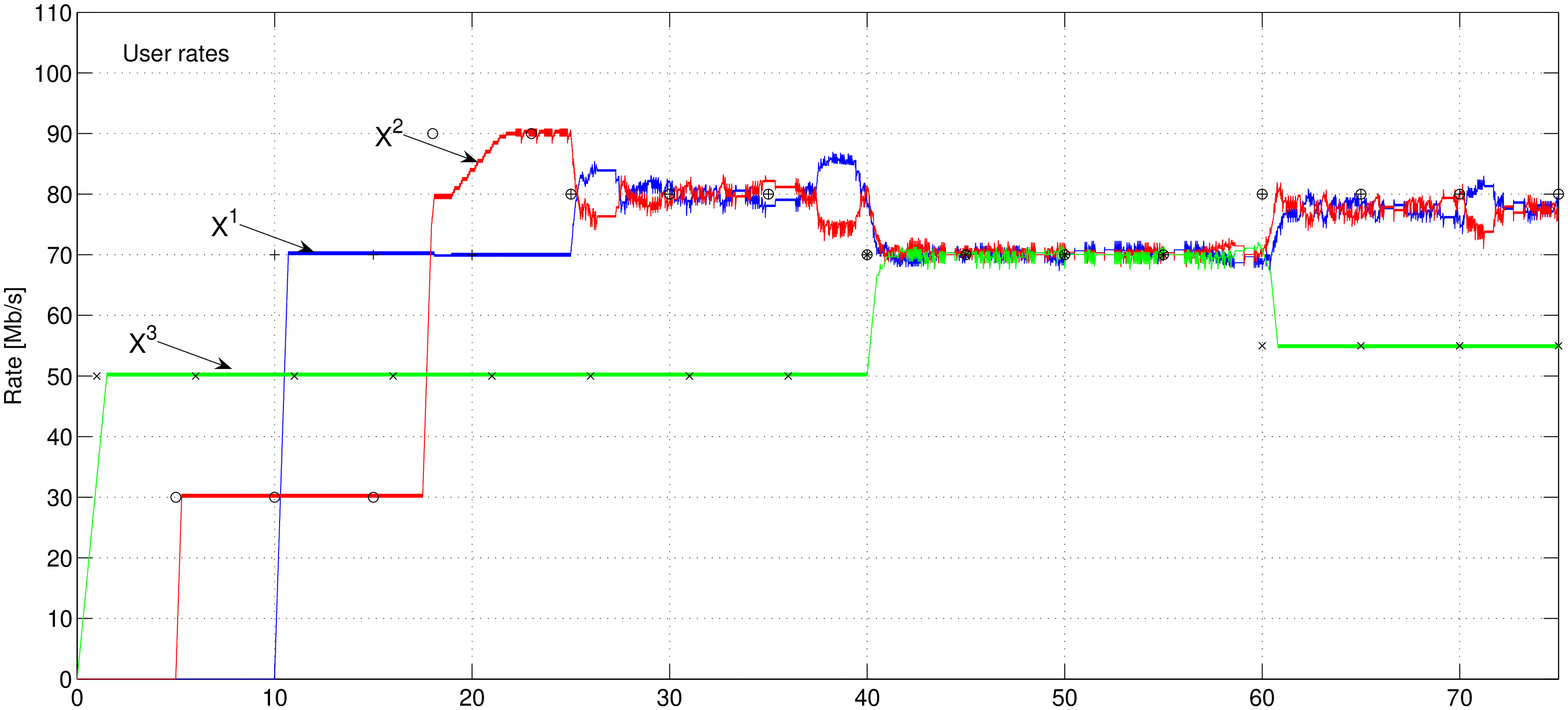} &
\includegraphics[width=0.5\textwidth,height=0.3\textwidth]{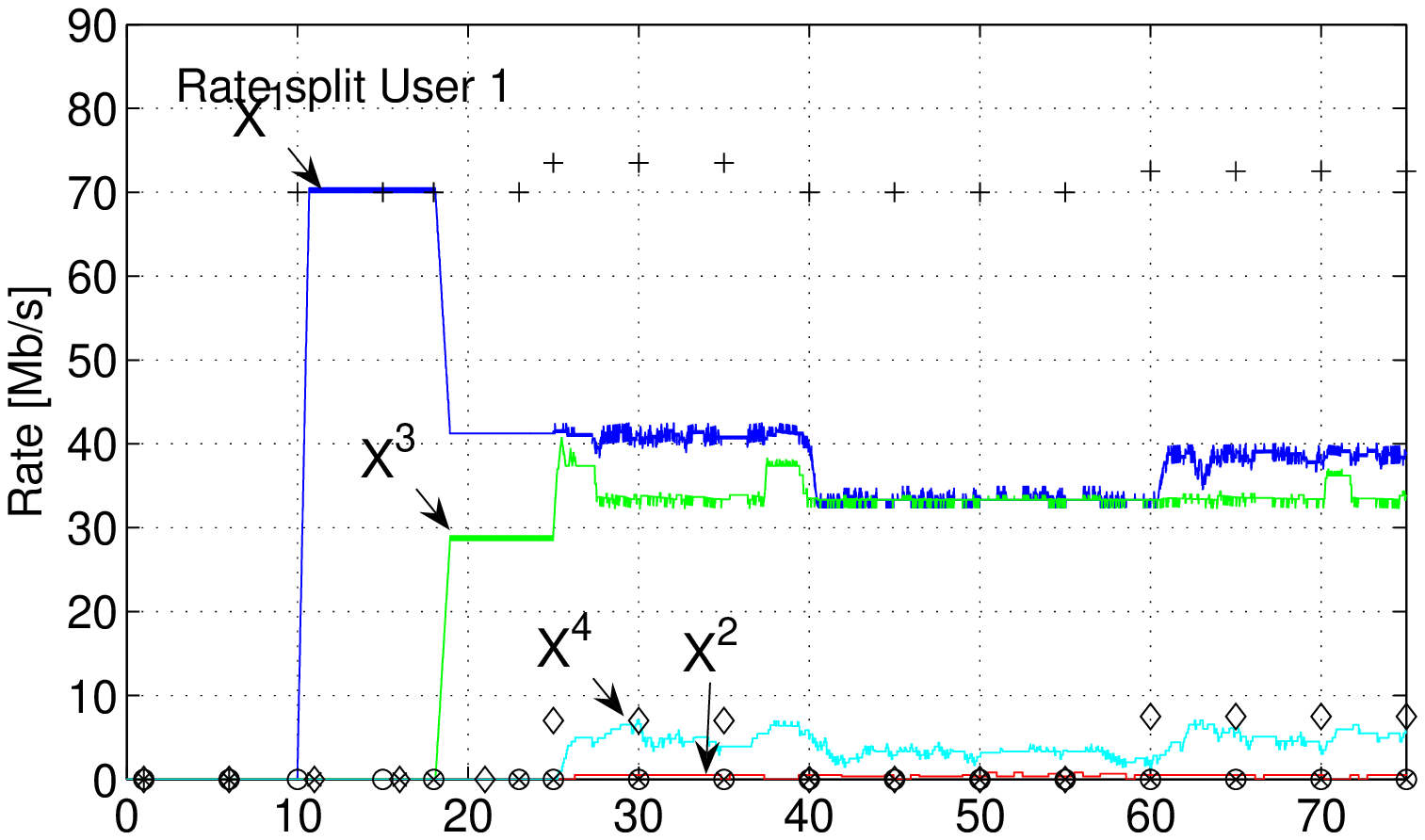}\\
\includegraphics[width=0.5\textwidth,height=0.3\textwidth]{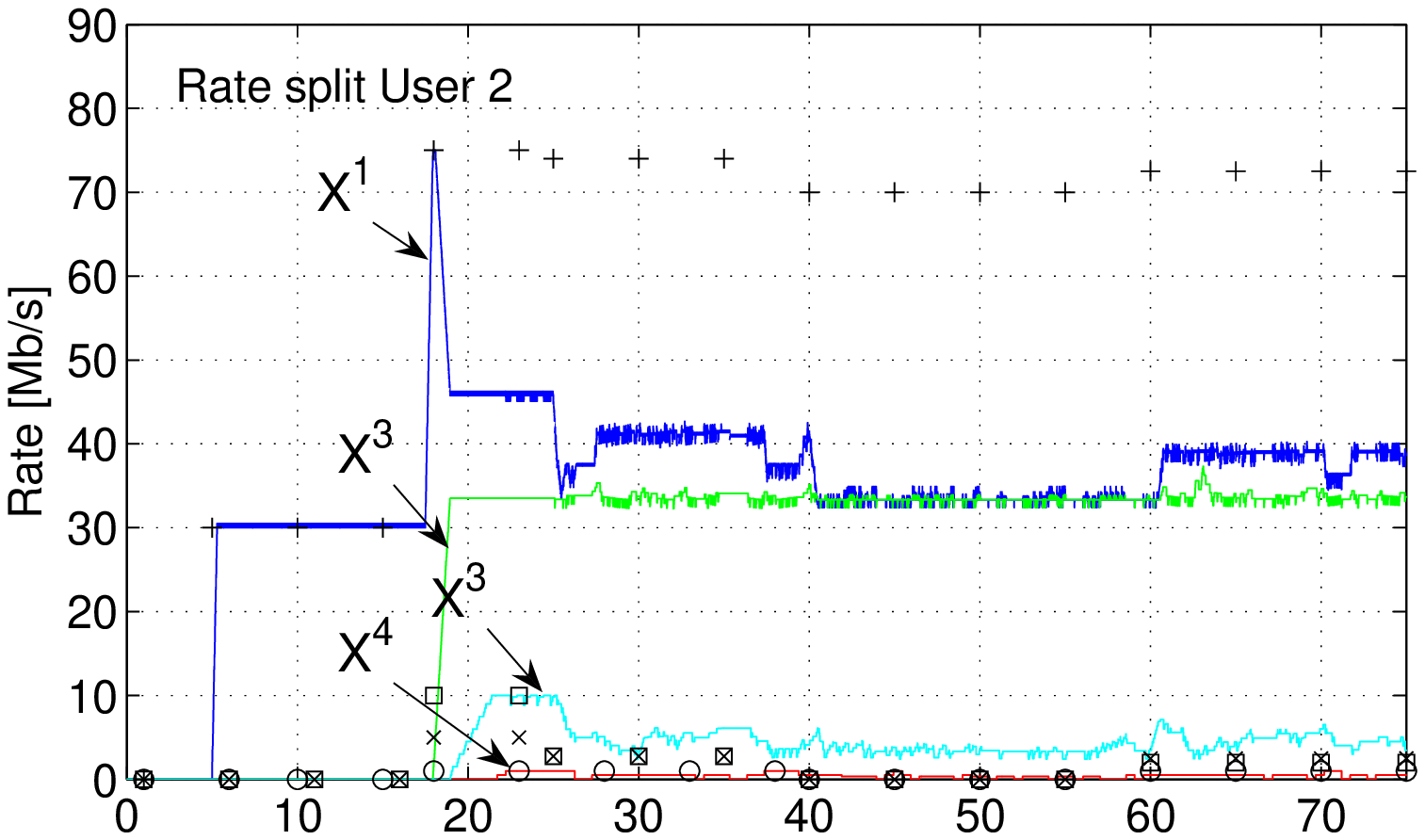} &
\includegraphics[width=0.5\textwidth,height=0.3\textwidth]{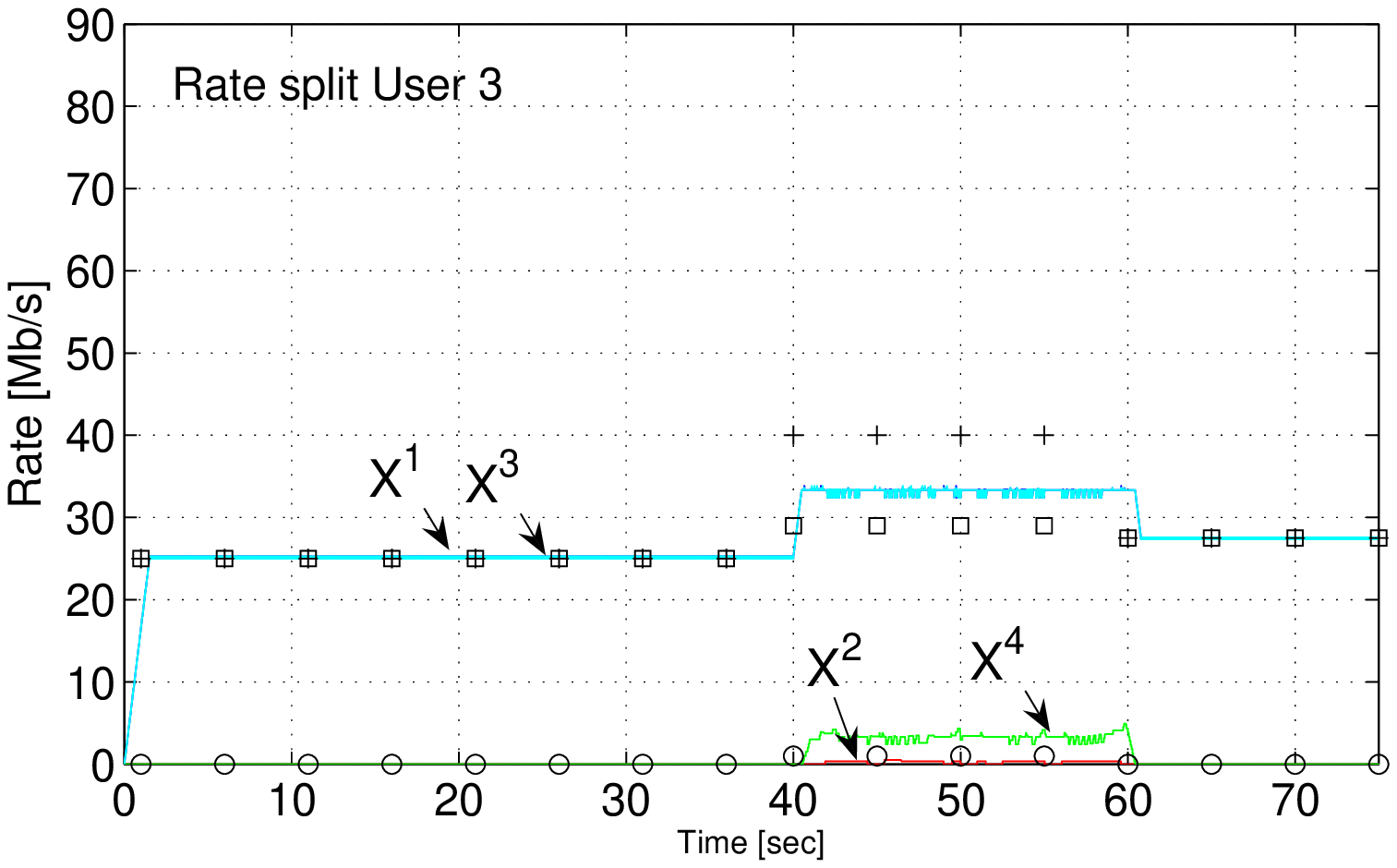}\\
\end{array}$
\end{center}
\caption{Time evolution of the rate allocation for MIRTO in a FA architecture. Points indicate the optimal allocation.}
 \label{fig:mirtoFA}
\end{figure}

Figure~\ref{fig:mirtoFD} shows the total flow rates and the flows splitting obtained by running MIRTO over a FD architectures. 
MIRTO rates are compared to those obtained by running the LP optimizer described in section~\ref{sec:largelp}.
As expected, total flow rates and rate splits over paths achieved by MIRTO follow those obtained by the LP. 
Fluctuations occur and are more pronounced when two or more flows are elastic. This is due to the probing nature of the 
algorithm that allows the global optimum attainment.\\
Figure~\ref{fig:mirtoFA}  shows flow rates obtained by running MIRTO over a FQ architecture and a comparison with the LP. 
Over this architecture only total flow rates follow the trend of the LP while the flow spitting is quite different from optimum. 
This confirms what stated in section~\ref{sec:arcprot} as FA breaks coordination among flows. 
In that way, flows cannot achieve rates larger than their fair rate over paths. 
This leads to a sub-optimal solution even if total flow rates are the same as before. 
In fact, they have been achieved with a larger network cost. As expected, with a FA architecture rate fluctuations 
are reduced and the algorithm converges in a shorter time.\\

\begin{figure}[htb]
\begin{center}
$\begin{array}{cc}
\includegraphics[width=0.5\textwidth,height=0.3\textwidth]{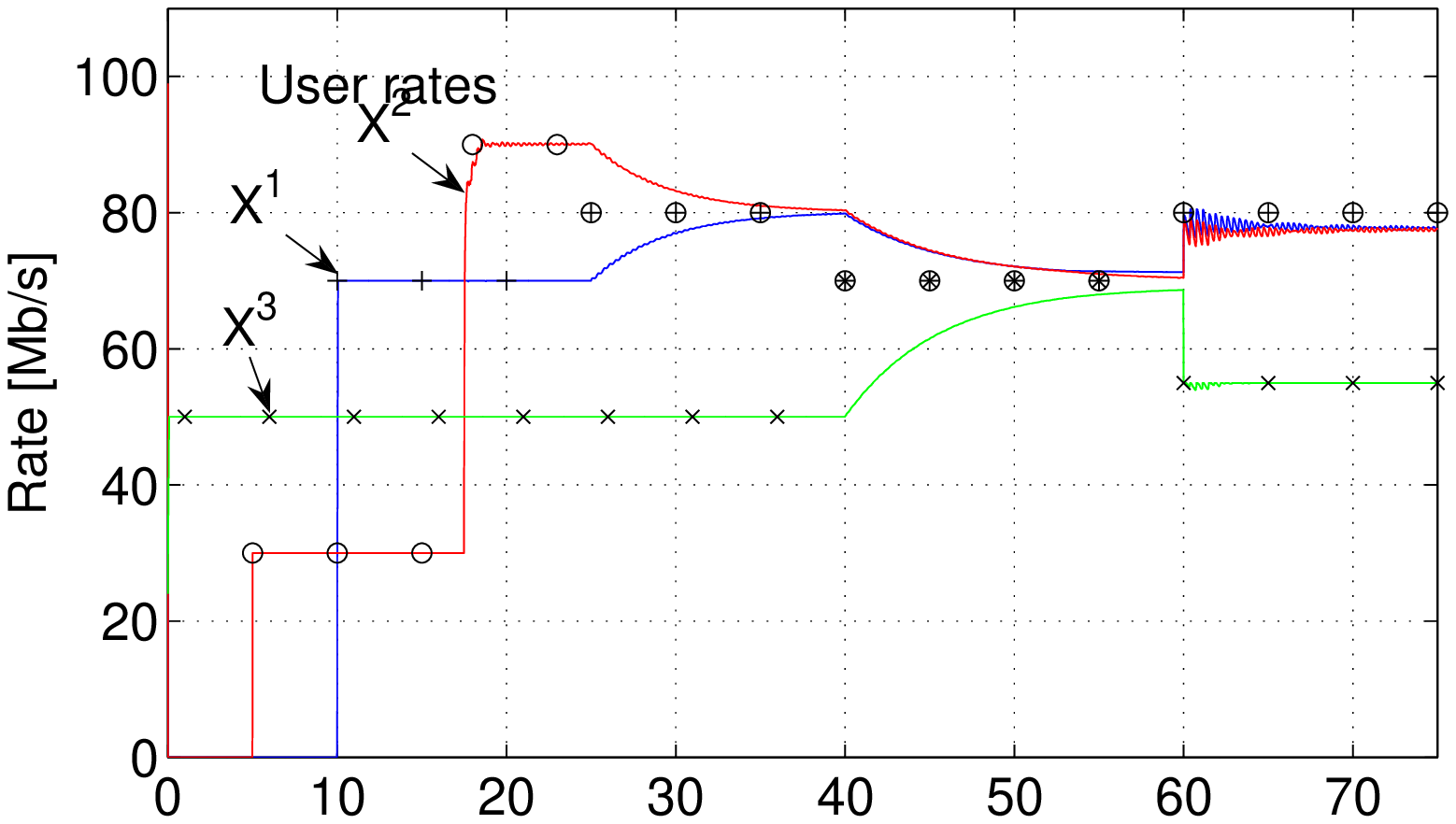} &
\includegraphics[width=0.5\textwidth,height=0.3\textwidth]{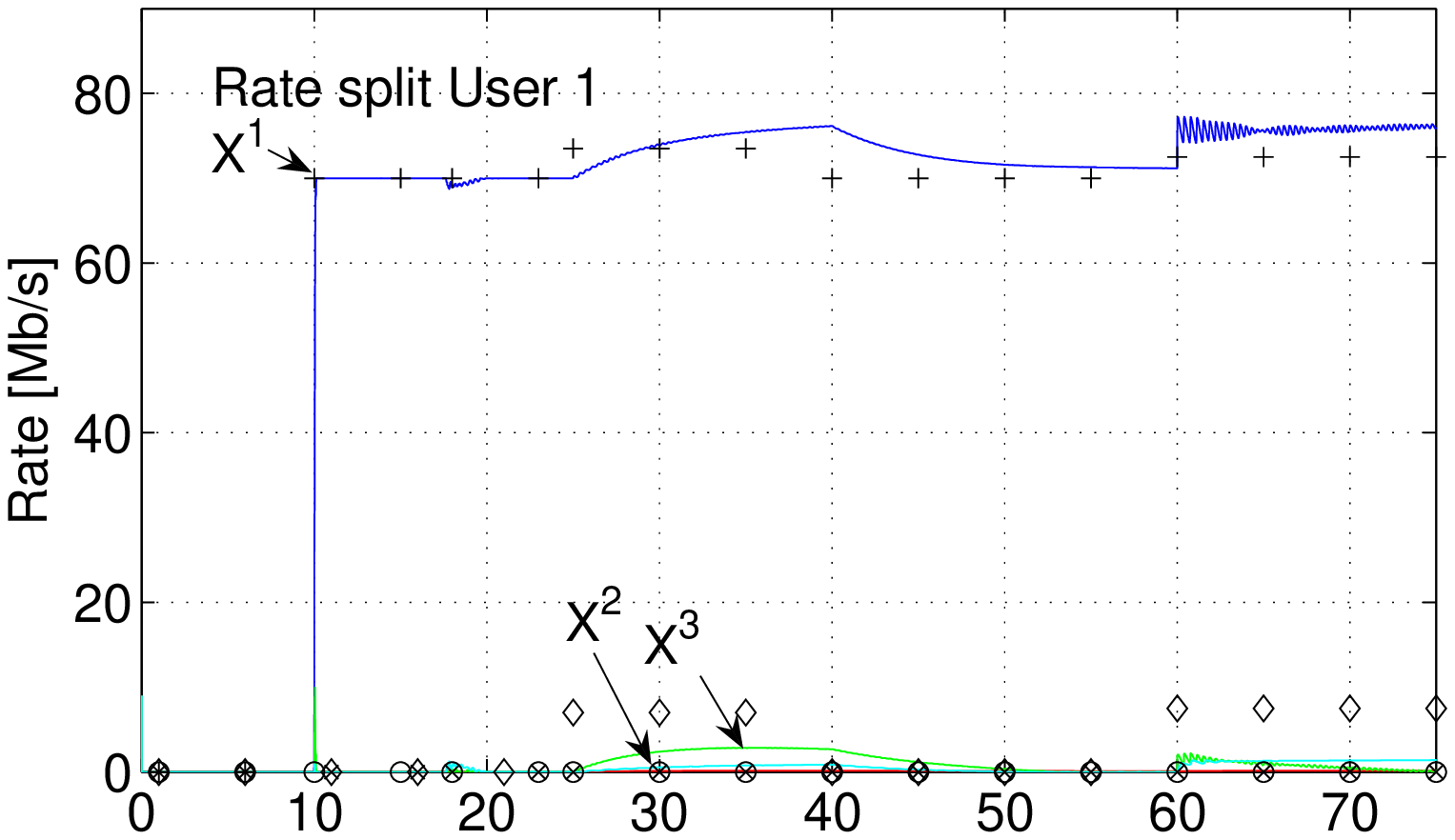}\\
\includegraphics[width=0.5\textwidth,height=0.3\textwidth]{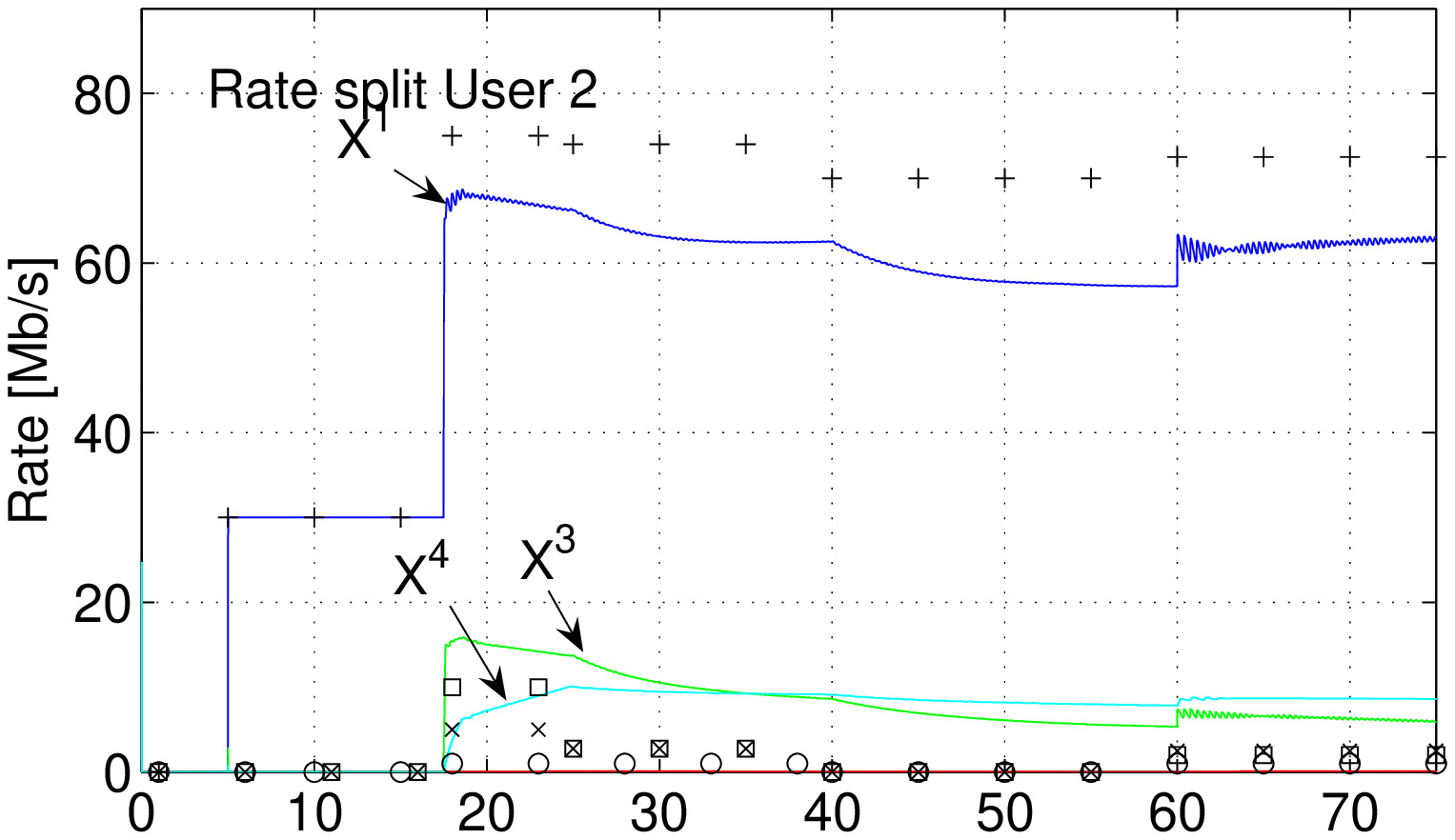} &
\includegraphics[width=0.5\textwidth,height=0.3\textwidth]{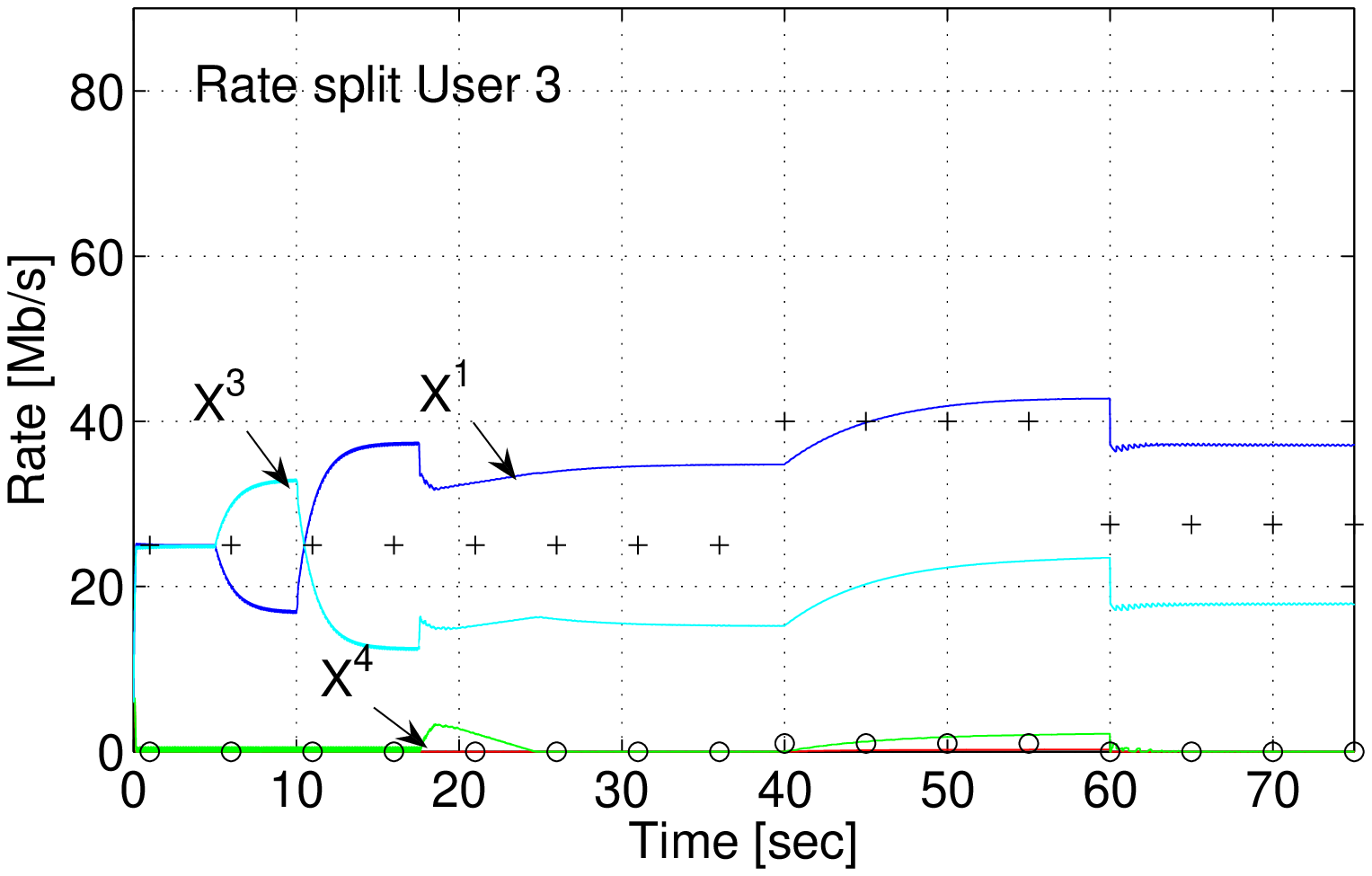}\\
\end{array}$
\end{center}
\caption{Time evolution of the rate allocation for TRUMP in a QD architecture. Points indicate the optimal allocation.}
 \label{fig:trump}
\end{figure}

Finally, Figure~\ref{fig:trump} reports performance of the TRUMP algorithm. 
TRUMP achieves optimal rates and splitting but it  takes long time to converge 
and some long term fluctuations occur. This is a consequence of the set of parameters we used.
With different values we would have seen a different behaviours with much faster 
convergence and no fluctuations. However the optimisation solution
would have been significantly different from that we want to achieve. 
Actually this is not a drawback of TRUMP as it has been designed in the context of intra-domain
TE to limit the link loads within the network, even though authors assume possible to 
implement such controller at end hosts. Unfortunately TRUMP does not allow to predict at which level of utilisation 
link can be set. Hence the choice of parameters might be very complex 
in networks with heterogeneous capacities.

We have tested TRUMP for different setup and different parameters and measured its good properties in 
term of fast convergence as the authors in \cite{RexfordCo07}.

This show the choice of parameters in such algorithms is a key 
point and could be quite complicated. Moreover it is strongly related to topology and 
could drive to very different solutions.

\section{Discussion and Conclusions}
The outcomes of this papers are multiple:
\begin{itemize}
\item we have studied congestion control and multi-path routing
in presence of demands with exogenous peak rates and we have found
that multiple routes can be likely exploited by higher rate flows.
\item in network scenario that are common in practice, coordinated
and uncoordinated multi-path routing perform the same.
\item in typical scenarios where coordination outperforms
un-coordination, multiple path does not gain much with respect to single
path routing. 
\item furthermore we propose a new multi-path optimal controller called MIRTO.
This controller is able to allocate max-min bandwidth among demands exploiting
path diversity.  The controller has been compared with other solutions and has proved
to be easy to be deployed in different architectures.
\end{itemize}
The above mentioned outcomes allow to positively rethink a flow aware architecture,
intrinsically uncoordinated, in presence of multi-path routing. 
Indeed multi-path controllers as MIRTO can be  effectively used in order to exploit path
diversity even when a flow aware network imposes fairness at a link base.
Task of the controller remains to dynamically define split ratios over the paths
as network conditions change. 

We suggest the use of multiple paths for a certain class of applications that are 
naturally robust to variability as adaptive video streaming, P2P file sharing and CDN 
as they can effectively exploit unused capacity within the network in both Internet
backbones and wireless mesh backhauling systems.
On the other hand, taking into account the overhead required by multiple routes transmissions,
this  should not be used for more conversational traffic as voice, web, 
or short transactions as mail or instant messaging.\\

\end{document}